\documentclass[journal,10pt,twocolumn,twoside]{IEEEtran}
\usepackage[noadjust]{cite}

\ifCLASSINFOpdf
   \usepackage[pdftex]{graphicx}
\else
\fi
\usepackage{amsmath,amssymb,amsfonts,amsthm}
\usepackage{array}

\ifCLASSOPTIONcompsoc
  \usepackage[caption=false,font=normalsize,labelfont=sf,textfont=sf]{subfig}
\else
  \usepackage[caption=false,font=footnotesize]{subfig}
\fi
\newtheorem{theorem}{Theorem}
\newtheorem{proposition}{Proposition}
\newtheorem{corollary}{Corollary}

\newtheorem*{assumption*}{Assumption}

\newtheorem*{problem*}{Problem}

\begin{document}
%
\title{Range-Only Localization in n-Dimensional Networks With Arbitrary Anchor Placement}
%
%
%

\author{Pedro P. V. Tecchio*,
        Nikolay Atanasov,
        and~George~J.~Pappas
\thanks{P. P. V. Tecchio and G. J. Pappas are with the Department of Electrical and Systems Engineering, University of Pennsylvania, Philadelphia,
PA, 19104 USA (e-mail: tecchio@seas.upenn.edu; pappasg@seas.upenn.edu).}
\thanks{N. Atanasov is with the Department of Electrical and Computer Engineering, University of California, San Diego, CA, 92093 USA e-mail: natanasov@eng.ucsd.edu.}}
\maketitle

\begin{abstract}
This paper considers node localization in static sensor networks using range-only measurements. Similar to state-of-the-art algorithms, such as ECHO and DILOC, we rely on barycentric coordinates of the nodes to transform the non-convex node localization problem into a linear system of equations. The main contribution of this paper is a simple closed-form expression for generalized barycentric coordinates, which extends existing algorithms from two to n dimensions and allows arbitrary anchor-node configurations. The result relies on a connection between the Cayley-Menger bi-determinants of subsets of n+1 neighbor nodes and the signed volume of the simplices defined by these neighbor nodes. Hence, for noise-free measurements, the proposed method computes the optimal sensor network embedding as the solution of a linear system with coefficients obtained from the generalized barycentric node coordinates. Using simulations, we provide comparisons with DILOC and Matlab's MDS implementation. We also show that it is possible to improve our algorithm run time using fewer subsets of neighbor nodes.  
\end{abstract}

\begin{IEEEkeywords}
Sensor Networks, Localization, Cayley-Menger bi-determinant.
\end{IEEEkeywords}


%
\IEEEpeerreviewmaketitle

\section{Introduction}
\label{sec:intro}

\IEEEPARstart{L}{ocalization} problems are fundamental to a multitude of applications and everyday scenarios, including deployment of robots and drones, autonomous vehicular systems and static sensor networks. In the later case, the location of each sensor is usually necessary to correctly analyze, interpret and correlate all measurements being made.   

In general, localization problems in static sensor networks have a single objective: based on available measurements and known information about the surrounding space, provide locations of one or more sensor nodes. Multiple methodologies have been proposed. Some utilize range and bearing measurements between sensors \cite{NikolayJointEstimation}; while others utilize only bearings \cite{ShamesBearingOnly}, or only ranges \cite{ThomasTrilateration}, \cite{KhanDiloc} and \cite{DiaoECHO}.


These methods can also be classified depending on how the underlining algorithm computes the location of each sensor in the network. If all sensor nodes in the network send their information to one node, responsible for computing their locations, the method is called centralized. If each sensor node is responsible for computing its own location by exchanging information within its local neighbors, it is called distributed.   

In \cite{PryanthaN} and \cite{DiaoECHO}, localization methods are also classified as either sequential or concurrent. Sequential methods start from sensor nodes with known locations, called {\it anchors}, and compute location of other nodes individually or in groups. We call these latter nodes {\it unknown} throughout this paper. Concurrent methods start with an initial estimate of all node locations and finish when location estimates converge. At each iteration, nodes update their locations {by exchanging information and} using inter-node distance measurements with its neighbors. Trilateration methods, \cite{ThomasTrilateration}, are usual examples of sequential methods, while optimization like approaches such as MDS - Multi Dimensional Scalling, \cite{FrancoMDS} and \cite{KruskalMDS}, are examples of concurrent methods.

Trilateration is one of the most straight forward approaches to solve the range-only localization problem. Its approach involves solving sets of non-linear equations, in which measured distances between unknown nodes and anchor nodes must be equal to the Euclidean norm of their Cartesian coordinates. In the two-dimensional case, a generic example involves three anchor nodes $\{i,j,k\}$ and one unknown node $\{l\}$, such that the Trilateration equations become
\begin{equation}
\left\{
\begin{array}{lcl}
d(\mathbf{x}_l,\mathbf{x}_i) &=& ||\mathbf{x}_l - \mathbf{x}_i||_2\\
d(\mathbf{x}_l,\mathbf{x}_j) &=& ||\mathbf{x}_l - \mathbf{x}_j||_2\\
d(\mathbf{x}_l,\mathbf{x}_k) &=& ||\mathbf{x}_l - \mathbf{x}_k||_2
\end{array}
\right.
\end{equation}

In \cite{ThomasTrilateration}, Thomas et al. propose a modification of the Trilateration method using barycentric coordinates to localize a robot in two-dimensional space knowing the location of three points and the distances from those points to the robot. Although they propose the utilization of Cayley-Menger bi-determinants and determinants, also defined in \cite{blumenthal1970theory}, to compute all necessary coordinates, they emphasize a greater geometrical view of the problem. Relationships between barycentric coordinates, Cayley-Menger bi-determinants and geometrical notions such as angles, dot and cross products of vectors in two and three dimensional spaces are demonstrated.

In \cite{KhanDiloc}, Khan et al. proposed the Distributed Iterative LOCalization (DILOC) algorithm that can be classified as a range-only distributed concurrent method. This algorithm is defined in the $n$-dimensional Euclidean space, $\mathbb{R}^n$, and requires the following main assumptions: 1) there are $n+1$ anchor nodes; 2) all unknown nodes are placed inside the convex-hull of the anchor nodes; 3) for each unknown node, there is at least one subset of $n+1$ of its neighbors such that the former lies in the convex-hull of the latter.

If the sensor network satisfies the previous assumptions, DILOC can be initialized at each unknown node by computing its barycentric coordinate in relation to one of its subsets of neighbors given by assumption 3). Later, an iterative process starts using location estimates for each unknown node and true locations for anchor nodes. At each subsequent step, neighboring nodes exchange their location estimates; unknown nodes update their location estimates based on the convex-combination of their neighbors' estimates and its barycentric coordinates, while anchor nodes maintain their assigned locations.

This iterative process is proven to converge to the true solution for all nodes in \cite{KhanDiloc} by reorganizing all barycentric coordinates in matrix form, one row for each network node. The resulting matrix is right stochastic, which allows one to see the entire iterative process as an Absorbing Markov Chain which converges to the desired result. 

In \cite{DiaoECHO}, Diao et al. propose modifications to DILOC's algorithm relaxing its second and third previously specified assumptions, which they call Extended Computation scHeme of cOordiate (ECHO). Thus, ECHO enables arbitrary locations for anchor nodes in the network and allows the utilization of subsets of neighbors in which a node does not strictly lie inside the convex-hull of its neighbors.

When the latter condition occurs; though the barycentric coordinates of a node can never be all simultaneously negative, some will be negative or zero. In order to compute such coordinates, Diao et al. present a series of algorithms based on geometrical properties of the two dimensional case. They also propose the concept of generalized barycentric coordinates, as the average of all possible barycentric coordinates of a node. 

Similarly to DILOC, these coordinates can be arranged in matrix form, as a linear system in terms of anchor and unknown node locations. Conversely, ECHO's matrix may not be right stochastic, so one can not relate it to Markov Chain theory. Moreover, this linear system may have unstable eigenvalues which can cause problems  with iterative and distributed solution methods. In spite of that, Diao et al. show in \cite{DiaoECHO} a feasible distributed algorithm, as well as, proof that, in the two dimensional case, ``an entire sensor network is localizable if and only if every sensor node has at least three disjoint paths to the anchor nodes in the graph associated with the barycentric coordinate representation."

Unfortunately, ECHO is only defined for two dimensional Euclidean spaces, restricting its applicability from scenarios involving 3D ad hoc networks, such as the deployment of robots and drones. We propose a generalization of ECHO over $n$-dimensional Euclidean spaces, as well as, a more concise way to compute barycentric coordinates on any number of dimensions and node arrangements. Our algorithm was influenced by ideas presented in \cite{ThomasTrilateration}, but like ECHO, an increase of connectivity on the network will incur in an increase computational expense in order to compute the generalized barycentric coordinates from all possible combinations of $n+1$ neighbor subsets of each node. 

Our main contributions to the static sensor network node localization problem using barycentric coordinates are:
\begin{enumerate}
 \item Arbitrary anchor node placement among unknown nodes;
 \item Arbitrary n-dimensional Euclidean spaces allowed; 
 \item Extension of generalized barycentric coordinates using Cayley-Menger bi-determinants.
\end{enumerate} 

The solution we provide is centralized, but a method similar to the one proposed in \cite{DiaoECHO} can also be applied to compute all sensor node locations in a distributed form.

In the upcoming sections, we formally state the localization problem in sensor networks while providing the necessary concepts of graph theory. We define Cayley-Menger bi-determinants and determinants, and show how to use them in order to compute barycentric coordinates of nodes in $n$-dimensional Euclidean space, as well as their generalization similarly to \cite{DiaoECHO}. Finally, we show how to use these generalized barycentric coordinates to compute all unknown node coordinates.

%
%
%
%

\section{Problem Formulation}
\label{sec:problem}

\begin{figure}[!b]
\centering
\includegraphics[width=0.35\textwidth]{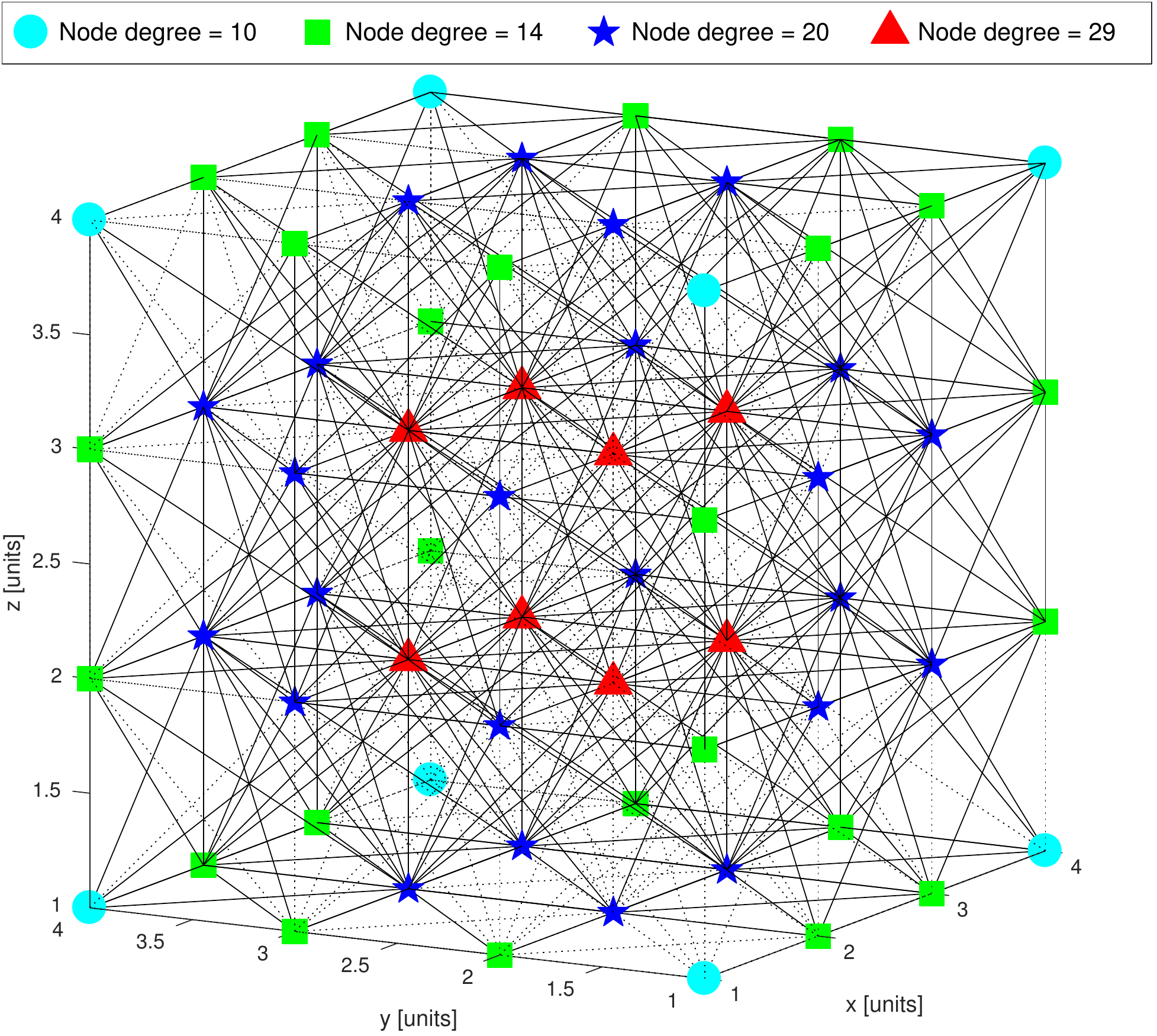}
\caption{Wireless node network example.}
\label{fig:ProblemNetworkExample}
\end{figure}

A static sensor network can be modeled as a graph $\mathcal{G} = \{\mathcal{V},\mathcal{E}\}$, with vertex set $\mathcal{V}$ and edge set $\mathcal{E}$. The vertex set contains unique labels for each sensor node in the network. Without loss of generality, we assume that labels are ordered and taken from the set of non-zero natural numbers, $\mathcal{V} \subset \mathbb{N}_{\neq 0}$.

Let the location of node $i \in \mathcal{V}$ be specified by  Cartesian coordinates ${\bf x}_i \in \mathbb{R}^{n}$. If the number of nodes in the network is $m = |\mathcal{V}|$, then the set of node locations is given by $\mathcal{X} = \{\mathbf{x}_1,\mathbf{x}_2,\hdots,\mathbf{x}_m\}$. 

In general, edges exist whenever two nodes are able to communicate with one another. However, in this work, an edge exists whenever two nodes are able to communicate and measure their relative distance. Considering that each node $i \in \mathcal{V}$ has inter-node measuring and communication range given by ${\it r}_i \in \mathbb{R}_{>0}$, an edge between node pair $(i,j) \in V  \times V$ will exist, if and only if, its edge weight, $d({\bf x}_i,{\bf x}_j) = || {\bf x}_i - {\bf x}_j ||_2$, satisfies $d({\bf x}_i,{\bf x}_j) = d({\bf x}_j,{\bf x}_i)$ and $d({\bf x}_i,{\bf x}_j) \leq  \min ({\it r}_i, {\it r}_j)$. Therefore, the generated graph is undirected.

An example of such a network in 3D is shown in Fig. \ref{fig:ProblemNetworkExample}. In this example, all 64 nodes are arranged in a cubic lattice with a lattice constant of 1 unit. Inter-node measuring and communication ranges are 2 units, ${\it r}_i = 2$, $1 \leq i \leq m $. Fig. \ref{fig:ProblemNetworkExample} also shows the node degree\footnote{In graph theory, node degree refers to the number of edges a node has to other nodes in the network, which are its direct neighbors.} of each vertex in this network example. Node degree will influence the localizability of the network, as will be shown later. 

One can infer that a subset of nodes with known coordinates is required to compute all other node coordinates, that is, one needs location references. The subset of anchor node coordinates is specified as $\mathcal{X}_a \subset \mathcal{X}$ while the subset of unknown node coordinates is $\mathcal{X}_u \subset \mathcal{X}$ so that $\mathcal{X}_a \bigcup \mathcal{X}_u = \mathcal{X}$ and $\mathcal{X}_a \bigcap \mathcal{X}_u = \emptyset$.

Then, the range-only node localization in static sensor networks problem consists of estimating each unknown node coordinate using only range measurements between nodes.

\begin{problem*}
Given $\mathcal{X}_a \subset \mathcal{X}$ and $d({\bf x}_i,{\bf x}_j)$ for all $(i,j) \in \mathcal{E}$, estimate $\mathcal{X}_u \subset \mathcal{X}$.
\end{problem*}

Next, while making required extensions to the $n$-dimensional Euclidean space, we present the necessary concepts related to Cayley-Menger bi-determinants, barycentric coordinates and its generalization, as well as, the methodology to compute unknown node locations using these concepts from true anchor node locations and range measurements.
\section{The Barycentric Coordinates Approach}
\label{sec:TheBarycentricApproach}

\subsection{Cayley-Menger bi-determinants}
Cayley-Menger determinants provide a relation between Euclidean distances between points in space and the signed volume of the simplex formed by such points. \cite{KhanDiloc} and \cite{DiaoECHO} use the absolute value of these signed volumes in order to compute the barycentric coordinates of nodes in sensor networks from their noiseless range measures. 

Similarly to \cite{ThomasTrilateration}, we introduce the concept of Cayley-Menger bi-determinants. While, Cayley-Menger determinants operate on one set of points, bi-determinants operate on two sets of points; providing a relation between the product of volumes of each set and the Euclidean distances between points of different sets. 

We define both types of Cayley-Menger determinants following Blumenthal's work in \cite{blumenthal1970theory}.

Let two sets of $n+1$ points, $\mathcal{X} = \{{\bf x}_{0}, \hdots, {\bf x}_{n}\}$ and $\mathcal{Y} = \{{\bf y}_{0}, \hdots, {\bf y}_{n}\}$, be defined by their Cartesian coordinates, such that ${\bf x_i} = [x_{1i}, \hdots, x_{ni}]^T \in \mathbb{R}^{n \times 1}$ and ${\bf y_i} = [y_{1i}, \hdots, y_{ni}]^T \in \mathbb{R}^{n  \times 1}$ for $0 \leq i \leq n$. Then, the Cayley-Menger bi-determinant of $\mathcal{X}$ and $\mathcal{Y}$ is defined as follows:
\begin{equation}
\begin{array}{l}
D({\bf x}_0, \hdots, {\bf x}_{n};{\bf y}_0, \hdots, {\bf y}_{n}) = \\
2\left(-\frac{1}{2}\right)^{n+1} 
\left|
\begin{smallmatrix}
0 & 1                   & 1                   & \hdots & 1 \\
1 & d({\bf x}_0, {\bf y}_0)^2 & d({\bf x}_0, {\bf y}_1)^2 & \hdots & d({\bf x}_{0}, {\bf y}_{n})^2\\
1 & d({\bf x}_1, {\bf y}_0)^2 & d({\bf x}_1, {\bf y}_1)^2 & \hdots & d({\bf x}_{1}, {\bf y}_{n})^2\\
\vdots & \vdots         & \vdots              & \ddots & \vdots                 \\
1 & d({\bf x}_{n}, {\bf y}_0)^2 & d({\bf x}_{n}, {\bf y}_1)^2 & \hdots & d({\bf x}_{n}, {\bf y}_{n})^2\\
\end{smallmatrix}
\right|.
\end{array}
\label{eq:CayleyMengerBiDet}
\end{equation}

Where, $d({\bf x}_i, {\bf y}_j)^2 = ||{\bf x}_i - {\bf y}_j||^2 = ({\bf x}_i-{\bf y}_j)^T({\bf x}_i-{\bf y}_j)$, for all $0 \leq i,j \leq n$.

The Cayley-Menger determinant of $\mathcal{X}$ can be defined from the bi-determinant as follows: 
\begin{equation}
D({\bf x}_0, \hdots, {\bf x}_n) = 
D({\bf x}_0, \hdots, {\bf x}_n;{\bf x}_0, \hdots, {\bf x}_n).
\label{eq:CayleyMengerDet}
\end{equation}

As one can infer from equation \eqref{eq:CayleyMengerDet}, the Cayley-Menger determinant is a specific case of the more general bi-determinant. Next, we formally specify the relationship between the signed volumes of sets of points in $n$-dimensional Euclidean space and their Cayley-Menger bi-determinant.

\begin{proposition}
\label{prop:CayleyMengerAndVolume}
The Cayley-Menger bi-determinant of two sets of $n+1$ points, $\mathcal{X} = \{{\bf x}_{0}, \hdots, {\bf x}_{n}\}$ and $\mathcal{Y} = \{{\bf y}_{0}, \hdots, {\bf y}_{n}\}$, in $\mathbb{R}^n$ is related to the products of the signed volumes of each independent set, by 
\begin{equation}
D({\bf x}_0, \hdots, {\bf x}_n; {\bf y}_0, \hdots, {\bf y}_n) = 
(n!)^2\text{Vol}(\mathcal{X}) \text{ }\text{Vol}(\mathcal{Y}).
\end{equation}
\begin{proof}
See Appendix \ref{app:CayleyMengerBidet}.
\end{proof} 
\end{proposition}

It is very important to notice that determinants in general are alternating forms, so the order in which its elements are arranged is essential to its correct computation. 

The following sections will show how we leverage Proposition \ref{prop:CayleyMengerAndVolume} in order to compute barycentric coordinates of points inside and outside the convex-hull of its $n+1$ neighbors, which allow us to obtain similar, but more concise results than \cite{DiaoECHO}.

\subsection{Computing Barycentric Coordinates}
\label{sec:ComputingBarycentricCoordinates}

Barycentric coordinates are usually defined using concepts of points, affine spaces and affine frames. An affine space $\mathbf{X}$ is defined by a collection of points, a vector space and a function. An affine frame is a set of points in an affine space with origin ${\bf x}_0$, $\{{\bf x}_i\}_{i = 0,1,...,n}$, such that vectors $\{\overrightarrow{{\bf x}_0{\bf x}_1},\overrightarrow{{\bf x}_0{\bf x}_2},\cdots,\overrightarrow{{\bf x}_0{\bf x}_n}\}$ are linearly independent, i.e. they form a base for the embedded vector space $\overrightarrow{\mathbf{X}}$. By taking a field $\mathbf{K}$ such as $\mathbb{R}$, one can define barycentric coordinates as follows:

\begin{proposition}[{\cite[Prop.3.6.2]{berger2009geometry}}]
Let $\{{\bf x}_i\}_{i = 0,1,...,n}$ be a frame for an affine space $\mathbf{X}$. For any (point) ${\bf x} \in \mathbf{X}$ there exist $\lambda_i \in \mathbf{K}$, $0 \leq i \leq n$, such that $\sum_i \lambda_i = 1$ and ${\bf x} = \sum_i \lambda_i {\bf x}_i$. The scalars $\lambda_i$ are uniquely defined by this property and are called barycentric coordinates of ${\bf x}$ in the frame $\{{\bf x}_i\}_{i = 0,1,...,n}$.
\label{prop:AffineFrame}
\end{proposition}

From proposition \ref{prop:AffineFrame}, one can infer that in order to localize the unknown nodes in the network, one needs to know the locations of $n+1$ nodes in an $n$-dimensional space. Moreover, the $n+1$ anchor nodes must form an affine frame for the inherent affine space.

Suppose that our set of $n+1$ anchor node locations, represented by their Cartesian coordinates $\mathcal{X}_a = \{\mathbf{x}_{a_0}, \mathbf{x}_{a_1}, \cdots, \mathbf{x}_{a_n}\}$, form a frame for an affine space $\mathbf{X}$. We can compute the barycentric coordinates ${\bf \lambda}$ for any unknown node ${\bf x} \in \mathbf{X}$ by solving the following linear system, where $ X = [\mathbf{x}_{a_0} \mathbf{x}_{a_1} \cdots \mathbf{x}_{a_n}]$,
\begin{equation}
\begin{array}{llcl}
{\bf x} = \sum_i \lambda_i {\bf x_{a_i}}, &
\sum_i \lambda_i = 1 &
\Leftrightarrow &
\begin{bmatrix}
X \\ {\bf 1}^T
\end{bmatrix}
{\bf \lambda}
=
\begin{bmatrix}
{\bf x} \\ 1
\end{bmatrix}.
\end{array}
\label{eq:BarycentricCoordinatesDefinition}
\end{equation}

Because our set of anchor nodes form a frame in an affine space, the solution is unique by Proposition \ref{prop:AffineFrame}. Then,
$V_X =
\begin{bmatrix}
X \\ {\bf 1}^T
\end{bmatrix} 
$ has non-zero determinant. Using Cramer's Rule, we can compute each barycentric coordinate, $\lambda_i$, as follows
\begin{equation}
\lambda_i = 
\frac{
 \begin{vmatrix}
 X_i \\ {\bf 1}^T
 \end{vmatrix}}{
 \begin{vmatrix}
 X \\ {\bf 1}^T
 \end{vmatrix}
}
\end{equation}
where, $X_i$ is the matrix obtained from $X$ by replacing its $i^{th}$ column with ${\bf x}$.

The volume of the sets of points $\mathcal{X}_a$ and $\mathcal{X}_{a_i}$, where the $i^{th}$ point is replaced by ${\bf x}$, are given by
\begin{equation}
\begin{array}{lclclcl}
\text{Vol}(\mathcal{X}_a) 
&=&
\frac{1}{n!} \begin{vmatrix} X \\ {\bf 1}^T \end{vmatrix}, 
& & 
\text{Vol}(\mathcal{X}_{a_i})
&=& \frac{1}{n!} \begin{vmatrix} X_i \\ {\bf 1}^T \end{vmatrix}
\end{array}.
\end{equation}

Therefore, each barycentric coordinates, $\lambda_i$, can be computed in terms of these volumes by
\begin{equation}
\lambda_i = 
\frac{
 \begin{vmatrix}
 X_i \\ {\bf 1}^T
 \end{vmatrix}}{
 \begin{vmatrix}
 X \\ {\bf 1}^T
 \end{vmatrix}
}
=
\frac{(n!)\text{Vol}(\mathcal{X}_{a_i})}{(n!)\text{Vol}(\mathcal{X}_a)}. 
\end{equation}

Multiplying the right side by $ \frac{(n!)\text{Vol}(\mathcal{X}_a)}{(n!)\text{Vol}(\mathcal{X}_a)}$ does not change its value, but permits us to conclude that the barycentric coordinates, $\lambda_i$ for all $0\leq i\leq n$, can be computed using Cayley-Menger bi-determinants and determinants.

\begin{equation}
\lambda_i = \frac{D({\bf x_0,\hdots,x_n};{\bf x_0, \hdots, x_{i-1},x, x_{i+1}, \hdots, x_n})}{D({\bf x_0,\hdots,x_n})}
\label{eq:BarycentricCoordinates}
\end{equation}

We emphasize that equation \eqref{eq:BarycentricCoordinates} provides a method to compute barycentric coordinates of any point in an affine space based on any set of points that form an affine frame for that space. Moreover, if one maintains the ordering of points throughout all determinants, the signs of each barycentric coordinate will be correctly computed.

Notice that each point's barycentric coordinate sign is associated to the overall position of this point in relation to the ones forming the affine frame used in its computation. Any point strictly inside the convex-hull of the affine frame, will have strictly positive coordinates. Otherwise, it will have at least one zero or negative coordinate. It is impossible to have all non-zero barycentric coordinates with a negative sign. Moreover, the order in which the different signs are given depends on the partial ordering of the $n+1$ hyper-planes generated on the $n$-dimensional space.

It is important to notice that if we consider each node as a point in an affine space, then allowing nodes to be outside of the convex-hull of its neighbors, is equivalent of having zero or negative barycentric coordinates. This constitutes the main difference between DILOC's and ECHO's algorithm, \cite{KhanDiloc} and \cite{DiaoECHO} respectively.

\begin{figure}[!t]
\centering
\includegraphics[width=0.45\textwidth]{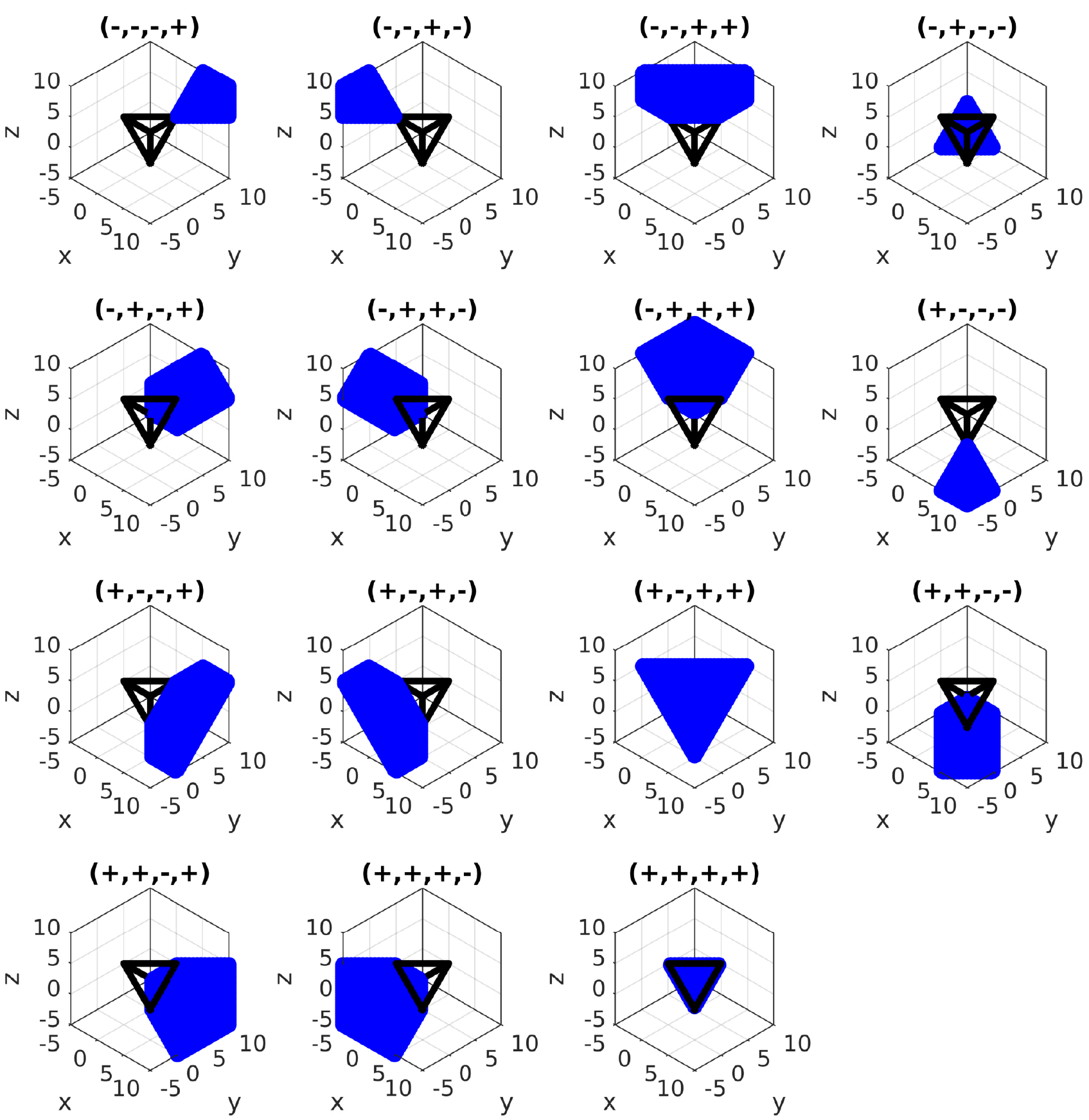}
\caption{Regions of different Barycentric Coordinate signs.}
\label{fig:GeneralizedBaryCoord}
\end{figure}

Fig. \ref{fig:GeneralizedBaryCoord} shows an example in three-dimensional space of the $2^{n+1}-1$ possible regions where the barycentric coordinate signs are strictly positive or negative. The convex-hull of the set of $n+1$ points forming the affine frame is depicted by having all its edges blacked in the figure.

\subsection{Generalizing Barycentric Coordinates}
We developed equation \eqref{eq:BarycentricCoordinates} by using the anchor nodes as an affine frame, but any set of $n+1$ nodes that form an affine frame for the same space can be used in order to compute different barycentric coordinates of the same node of interest, provided that one has all required range measurements between nodes. This concept was utilized in \cite{DiaoECHO} in order to create the Generalized Barycentric Coordinates, which we extend next for the $n$-dimensional case. 

Let $\mathcal{N}_l$ be the index set of neighbors of node $l$, defined as follows
\begin{equation}
\small{
\mathcal{N}_l = \left\{j \in \{1,2,\hdots,m \} \setminus \{l\} |\hspace{.5mm} ||\mathbf{x}_l-\mathbf{x}_j||_2 \leq \min{(r_l,r_j)}\right\}.}
\label{eq:Neighbors}
\end{equation}

Let $\mathcal{I}_{l}$ be a family of sets of $n+1$ indexes given by the combination without repetition of members of $\mathcal{N}_l$, which are also neighbors from one another. 

\begin{equation}
\mathcal{I}_{l} = \{
 \mathcal{V}_{J} \in \mathcal{N}_l^{\times n+1} | \mathcal{J} = \{\mathcal{V}_{J}, \mathcal{E}_{J}\}, \mathcal{E}_{J} \subset \mathcal{E} \text{ and } \mathcal{J} \in \mathbb{K}_{n+1} \}
\label{eq:FamilySets}
\end{equation}
Where, $\mathcal{J}$ is a subgraph of our network graph $\mathcal{G}$ and $\mathbb{K}_{n+1}$ is the set of complete graphs with $n+1$ vertices.

Therefore, the cardinality of $\mathcal{I}_{l}$ is $|\mathcal{I}_{l}| \leq \begin{pmatrix}
|\mathcal{N}_{l}| \\
n+1
\end{pmatrix}
$.

For each node $l$, we define sets of $n+1$ points in $\mathbb{R}^{n}$, $\mathcal{X}_{l_i}$, $1 \leq i \leq |\mathcal{I}_{l}|$, such that 
$$\mathcal{X}_{l_i} = \{\mathbf{x}_{\mathcal{I}_{l_{i1}}}, \mathbf{x}_{\mathcal{I}_{l_{i2}}}, \cdots, \mathbf{x}_{\mathcal{I}_{l_{i n+1}}}\},$$
and sets of $n+1$ points in the same space, $\mathcal{Y}_{l_{ij}}$, $1 \leq j \leq n+1$, such that 
$$\mathcal{Y}_{l_{ij}} = \{\mathbf{x}_{\mathcal{I}_{l_{i1}}}, \mathbf{x}_{\mathcal{I}_{l_{i2}}}, \cdots, \mathbf{x}_{\mathcal{I}_{l_{i j-1}}}, \mathbf{x}_{l}, \mathbf{x}_{\mathcal{I}_{l_{i j+1}}}, \cdots, \mathbf{x}_{\mathcal{I}_{l_{i n+1}}}\}.$$

Using the previously defined sets with the correct indexes applied to equation \eqref{eq:BarycentricCoordinates}, one can compute all possible barycentric coordinates for each node $l$. 

\begin{proposition}
\label{prop:BarycentricCoordinates}
For each node $l$ and each set $\mathcal{I}_{l_i}$, if $D(\mathcal{X}_{l_i}) \neq 0$, then the barycentric coordinates of node $l$, $\lambda_l$, with respect to its neighbors identified by the set $\mathcal{I}_{l_i}$ are
\begin{equation}
[\lambda_{l_{ij}}]_k = \left\{
\begin{array}{ll}
\dfrac{D(\mathcal{X}_{l_i};\mathcal{Y}_{l_{ij}})}{D(\mathcal{X}_{l_i})}, & \text{if } k = \mathcal{I}_{l_{ij}}\\
0, & \text{otherwise}\\
\end{array}
\right.. 
\end{equation}
\begin{proof}
 Follows from equation \eqref{eq:BarycentricCoordinates}, by taking the appropriate nodes and their indexes.
\end{proof}
\end{proposition}

If one arranges all computed barycentric coordinates for each node $l$ as previously specified in matrix form, with one row per combination, then the resulting matrix has $|\mathcal{I}_{l}|$ rows and $m$ columns. 

One possible way to utilize the previous result in Proposition \ref{prop:BarycentricCoordinates} is to concatenate all these matrices generating an overdetermined linear system. Another way is to compute the Generalized Barycentric Coordinates proposed by Diao et. al. in \cite{DiaoECHO}. These generalized coordinates are computed by averaging all possible barycentric coordinates for each node $l$ as stated in Proposition \ref{prop:GeneralizedBarycentricCoordinates}.

\begin{proposition}[\cite{DiaoECHO}]
\label{prop:GeneralizedBarycentricCoordinates}
For each node $l$, its generalized barycentric coordinate, $\boldsymbol{\lambda}_l \in \mathbb{R}^{m \times 1}$, can be computed as follows
\begin{equation}
[\boldsymbol{\lambda}_l]_j = \frac{1}{|\mathcal{I}_l|} \sum_{i = 1}^{|\mathcal{I}_l|}{\lambda_{l_{ij}}} \text{ , } 1 \leq j \leq m.
\label{eq:GeneralizedBaryCoord}
\end{equation}
\end{proposition}

One may wonder about the possibility of using a smaller subset of barycentric coordinates to compute a modified set of generalized coordinates. Is there a trade-off between accuracy and efficiency related to the number of used subsets? We seek to provide some insights via numerical simulations in section \ref{sec:Evaluation}.

As the generalized barycentric coordinates in equation \eqref{eq:GeneralizedBaryCoord} are computed through averaging, the property of summing to one is preserved. Moreover, one can use all $m$ generalized barycentric coordinates in order to compute the unknown nodes coordinates by simply solving a linear system as shown in the next section.

\subsection{Localizing the unknown nodes}

Finally, if one leverages the theory presented in the previous sections in a $n$-dimensional noiseless range-only static sensor network, one can utilize existing range measurements to compute generalized barycentric coordinates for each network node using Cayley-Menger bi-determinants as stated in Propositions \ref{prop:BarycentricCoordinates} and \ref{prop:GeneralizedBarycentricCoordinates}. 

\begin{theorem}
The unknown node locations, $X_u$, of the problem stated in section \ref{sec:problem} can be computed by solving the following linear system, whenever $(I - D)^{-1}$ exists:  
\begin{equation}
(I_{q\times q} - D_{q\times q}) \cdot X_{u_{q\times n}} = C_{q\times p} \cdot X_{a_{p\times n}}.
\label{eq:ErrorFreeSolution}
\end{equation}
Where, matrices $C \in \mathbb{R}^{q \times p}$ and $D \in \mathbb{R}^{q \times q}$ are given in equations \eqref{eq:LinearSystem} and \eqref{eq:LinearSystemBlockMatrices} and $X_a$ is the given location coordinates of the anchor nodes.
\begin{proof}
We known from Proposition \ref{prop:AffineFrame} that any unknown node can be localized in relation to neighboring nodes that form a frame in an affine space utilizing barycentric coordinates. Propositions \ref{prop:BarycentricCoordinates} and \ref{prop:GeneralizedBarycentricCoordinates} show us how to compute these barycentric coordinates. Thus, one can compute coordinates of each node as the linear combination of this node's generalized barycentric coordinates with the coordinates of all other nodes. 

Arranging these linear combinations in matrix form such that $G \in \mathbb{R}^{m\times m}$, $G^T = [\boldsymbol{\lambda}_1, \boldsymbol{\lambda}_2, \hdots, \boldsymbol{\lambda}_m]$, and all node Cartesian coordinates as $X \in \mathbb{R}^{m\times n}$, $X^T = [\mathbf{x}_1, \mathbf{x}_2, \hdots, \mathbf{x}_m]$. Then it is true that 
\begin{equation}
G \cdot X = X.
\label{eq:LinearSystem}
\end{equation}

Next, one can permute rows of $X$ in order to isolate anchor and unknown nodes, so that $X = [X_a; X_u]$. Applying the same permutation to the rows and columns of $G$ we can write
\begin{equation}
\begin{bmatrix}
A_{p\times p} & B_{p\times q}\\
C_{q\times p} & D_{q\times q}
\end{bmatrix}
\begin{bmatrix}
X_{a_{p\times n}}\\
X_{u_{q\times n}}
\end{bmatrix} = 
\begin{bmatrix}
X_{a_{p\times n}}\\
X_{u_{q\times n}}
\end{bmatrix},
\label{eq:LinearSystemBlockMatrices}
\end{equation}
where $p$ is the number of anchors, $n+1 \leq p < m$, and $q$ is the number of unknowns, $q = m - p$.

As we know the anchor node coordinates, then $X_a$ is defined and one can use equation \eqref{eq:ErrorFreeSolution} to find the unknown node coordinates $X_u$ exactly, if and only if the inverse of $I - D$ matrix exists.
\end{proof}
\label{theo:ProblemSolution}
\end{theorem}

The linear system in Theorem \ref{theo:ProblemSolution} has a unique solution, if and only if $(I_{q\times q} - D_{q\times q})^{-1}$ exists. In \cite{DiaoECHO},  Diao et al. provide necessary and sufficient conditions for its existence based on the number of disjoint paths from unknown nodes to anchor nodes in the two-dimensional case. As we provide a generalization of their problem to the $n$-dimensional case in Theorem \ref{theo:ProblemSolution}, we can state their result in Corollary \ref{cor:ECHO2DTheorem} with the necessary reference changes.

\begin{corollary}[Adapted from \cite{DiaoECHO}, Theorem 1]
A sensor network in $\mathbb{R}^2$ with generic configuration $[X_a^T X_u^T]$ is localizable using the barycentric coordinate representation by solving \eqref{eq:ErrorFreeSolution}, i.e. the matrix $I - D$ is nonsingular, if and only if every node to be localized has at least three disjoint paths from the set of anchor nodes in $\mathcal{G}_{\hat{G}}$. 
\label{cor:ECHO2DTheorem}
\end{corollary}  

Corollary \ref{cor:ECHO2DTheorem} makes reference to the undirected graph $\mathcal{G}_{\hat{G}} = (\mathcal{V}_{\hat{G}},\mathcal{E}_{\hat{G}})$. This undirected graph $\mathcal{G}_{\hat{G}}$ is constructed from our initial network representation graph $\mathcal{G}$. They both share the same set of vertices, $\mathcal{V}_{\hat{G}} = \mathcal{V}$, but possibly different edge sets. 

One may create an adjacency matrix from matrix $G$ defined in equation \eqref{eq:LinearSystem} assuming that elements of the latter are edge weights of some directed graph associated with the former. As it is assumed that edges in our network exists if and only if each node in a pair is able to communicate and measure their inter-node distances and there are no edges from one node to itself, then this adjacency matrix is symmetric. Lastly, Diao et al., in \cite{DiaoECHO}, further simplify this graph by modifying our original matrix $G$ to $\hat{G}$ given in equation \eqref{eq:MatrixAECHO}. Therefore, an edge $(i,j)$ exists in the edge set $\mathcal{E}_{\Hat{G}}$ of $\mathcal{G}_{\hat{G}}$, if and only if $[\hat{G}]_{ij} \neq 0$.

Notice, that $\hat{G}$ is constructed using block matrices from matrix $G$ in \eqref{eq:LinearSystemBlockMatrices} for the 2D case specifically. 
\begin{equation}
\hat{G} = \begin{bmatrix} I_{3\times 3} & 0_{3\times q} \\ C_{q\times 3} & D_{q\times q} \end{bmatrix}
\label{eq:MatrixAECHO}
\end{equation}

We hypothesize, without proof, that a similar result from Corollary \ref{cor:ECHO2DTheorem} exists for the n-dimensional case. In which case, $I - D$ would be nonsingular if and only if every node to be localized has at least $n+1$ disjoint paths from the set of anchor nodes in the respective undirected graph $\mathcal{G}_{\hat{G}}$. 
 
\subsection{Distributed algorithm} 
 
It is important to notice that similarly to what happens in \cite{DiaoECHO}, the linear system matrices associated with Theorem \ref{theo:ProblemSolution} may have eigenvalues with modulus greater than one, which can cause convergence problems if one tries to solve these systems in an iterative or distributed form. Because of that, Diao et al. provided a distributed method based on the Richardson iteration \cite{Richardson307}. This method can also be applied to the generic $n$-dimensional case presented here. 

In the following section, based on numerical simulations of our work, we provide comparisons with the standard DILOC algorithm \cite{KhanDiloc} and Matlab's MDS implementation, as well as, our algorithm complexity analysis and experimental run times.
\section{Evaluation}
\label{sec:Evaluation}

\subsection{Comparison with other algorithms}

\begin{figure}[!b]
  \centering
  \includegraphics[width=0.45\textwidth]{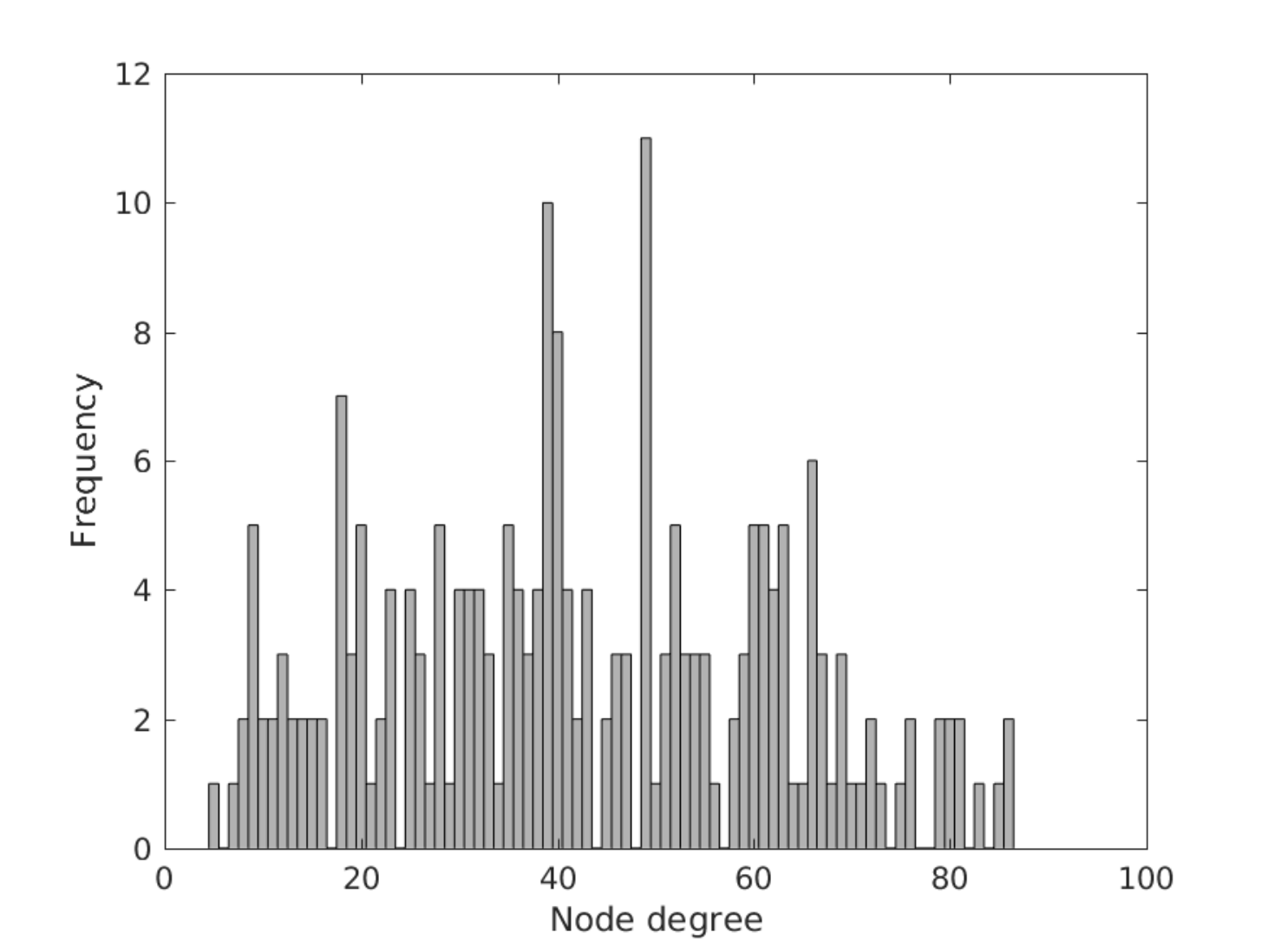}
  \caption{Node degree histogram for the 215 localized nodes in the static sensor network with maximum range radius of 3 units for each node.}
  \label{fig:ourNodeDegree}
\end{figure}

\begin{figure}[!b]
\centering
\subfloat[Our computed node coordinates.]{\includegraphics[width=0.4\textwidth]{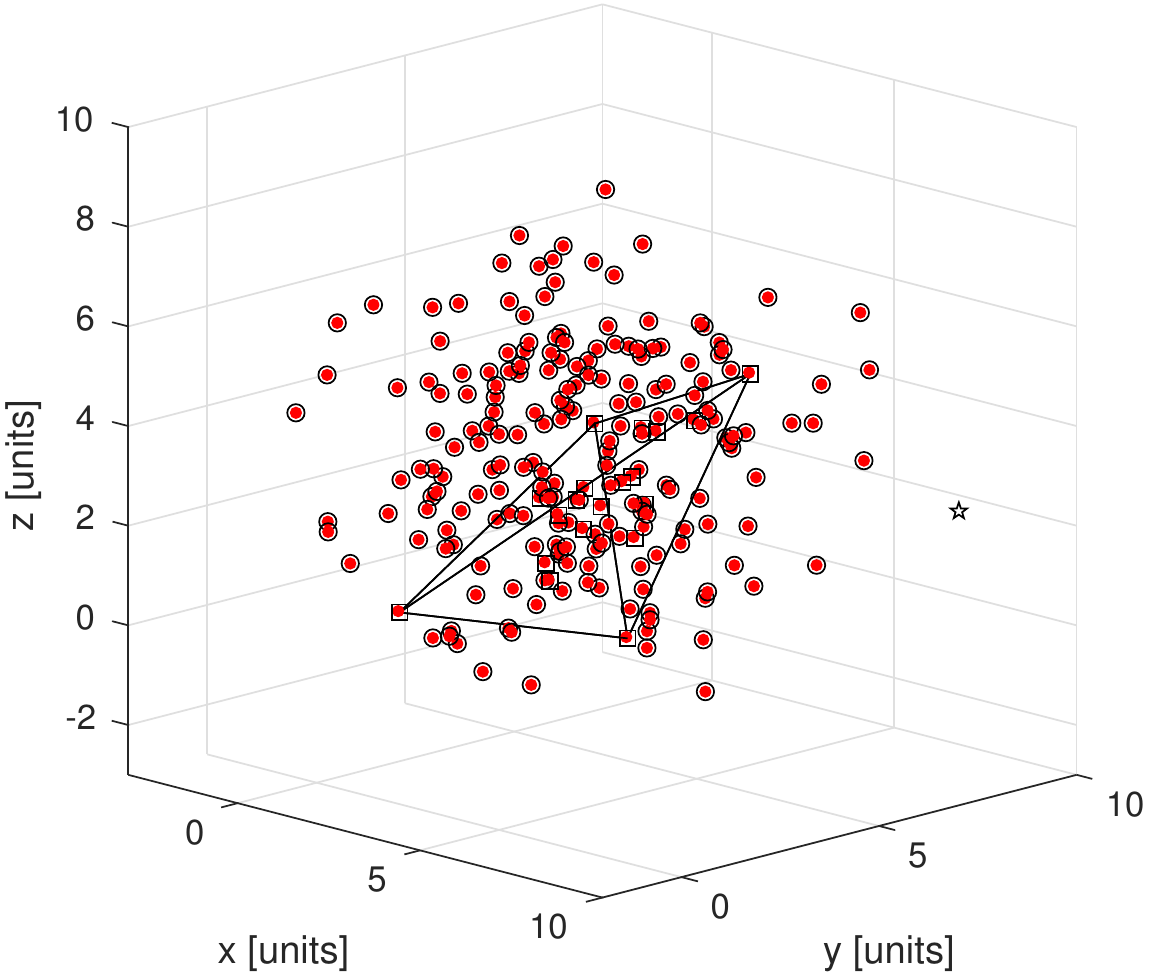}%
\label{fig:ourNodeCoord}}
\vfil
\subfloat[DILOC's computed node coordinates.]{\includegraphics[width=0.4\textwidth]{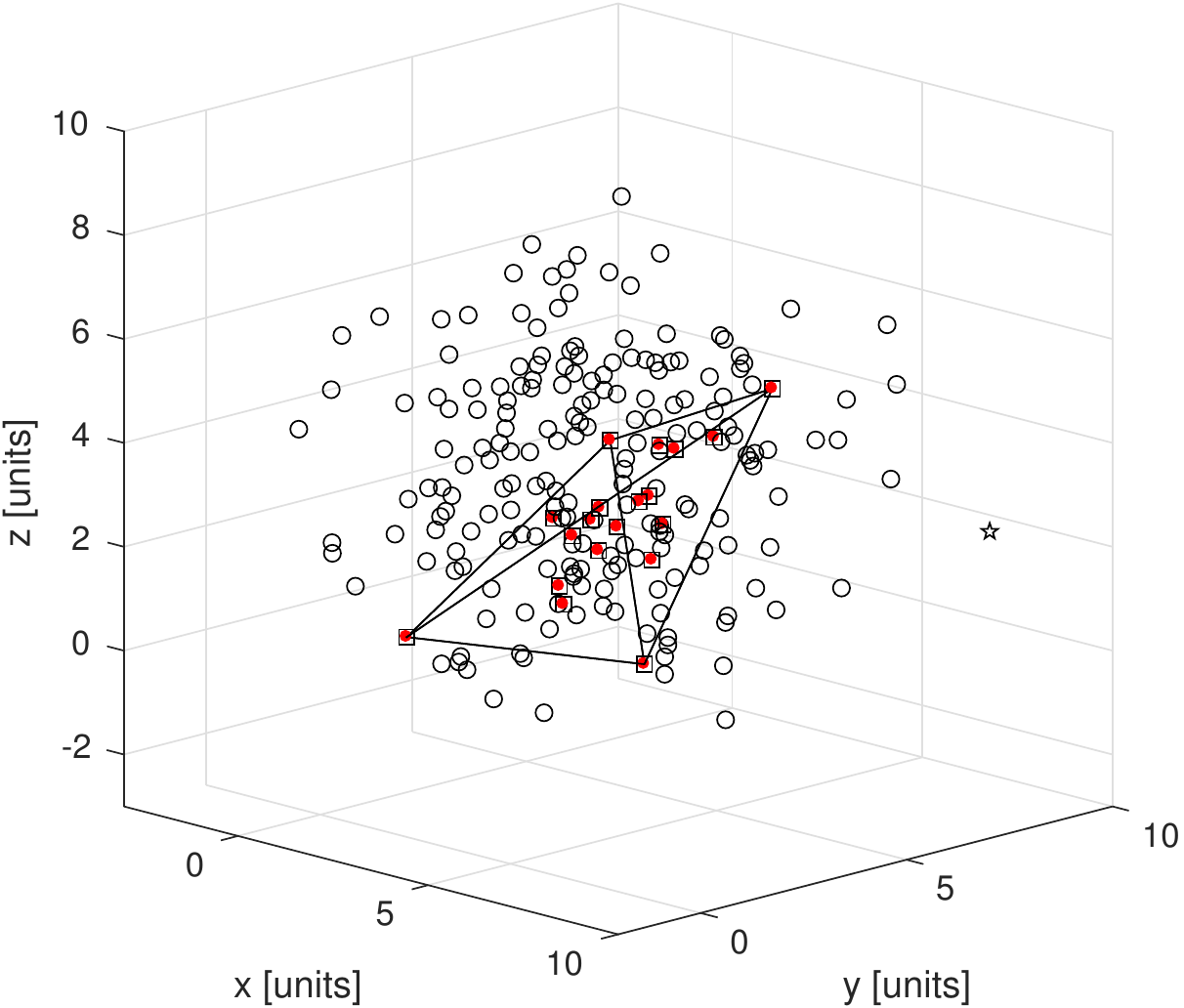}%
\label{fig:dilocNodeCoord}}
\vfil
\subfloat[MDS's computed node coordinates with the least stress.]{\includegraphics[width=0.4\textwidth]{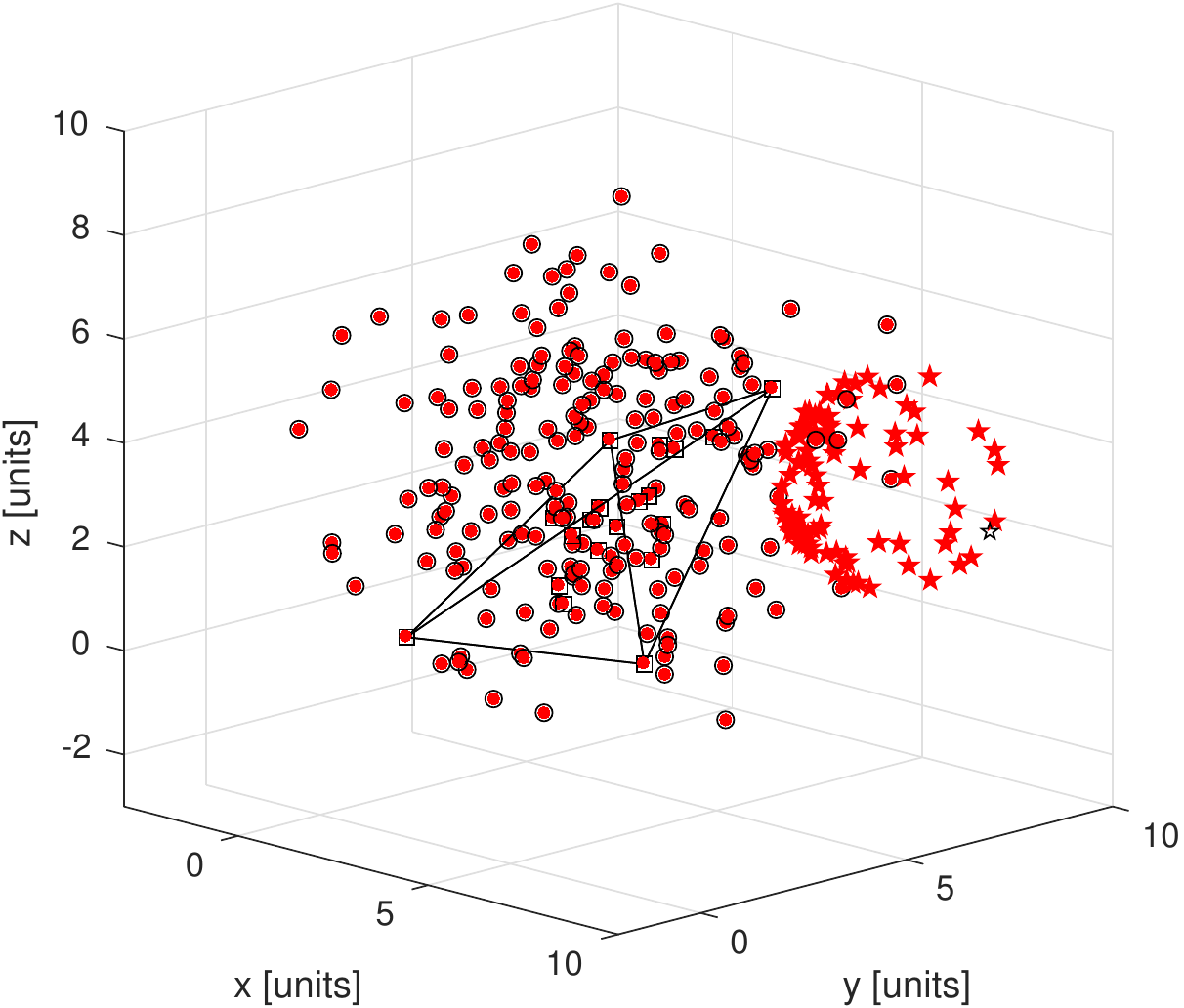}%
\label{fig:mdsNodeCoord}}
\caption{Static sensor network localization example with 216 nodes. Given the high number of edges in this network, they are not draw. Instead, we provide a histogram of its node degrees in Fig. \ref{fig:ourNodeDegree}. In these plots, the black frame represent edges of the anchor's convex-hull; black circles and squares are true node coordinates outside and inside the convex-hull respectively. Red dots are computed coordinates in each method, with the exception of one node which was not correctly localized by any algorithm. The latter has true coordinates marked by a black star, while red stars marks its MDS' computed values in which the stress criterion was inferior to 0.01.}
\label{fig:NodeCoord}
\end{figure}

In Fig. \ref{fig:NodeCoord}, we provide simulations of our proposed algorithm, the standard DILOC algorithm, \cite{KhanDiloc}, and Matlab's MDS implementation, $mdscale(.)$. The static sensor network from these examples was formed from a $6 \times 6 \times 6$ unit-spaced regular lattice in which independent Gaussian noise with zero mean and unit variance was added to each node coordinate. The maximum range threshold for each node $i$ was set up to $r_i = 3$ units. As the resulting number of edges is high, they were not draw in Fig. \ref{fig:NodeCoord}. Instead, we provide a histogram of node degrees for this network in Fig. \ref{fig:ourNodeDegree}.  Lastly, anchor nodes were chosen such that there were some nodes inside and outside their convex-hull.

%
%

From the 216 nodes in our example, 215 were correctly localized by the proposed method, as shown in Fig. \ref{fig:ourNodeCoord}, in which black circles and squares represent their true node coordinates, while red dots represent computed coordinates. The unlocalized node has its true coordinates represented by the sole black star.

As one can expect, DILOC's standard algorithm is able to localize all nodes inside the convex-hull of the anchor nodes as shown in Fig. \ref{fig:dilocNodeCoord}. Though, in order to accomplish that, 6 of the 15 unknown nodes inside the convex-hull had to use measurements greater than our proposed maximum range of 3 units. If one limits DILOC's range to 3 units, some nodes would not be localized, as these nodes would not have the necessary number of neighbors to compute their barycentric coordinates.

Matlab's MDS implementation, $mdscale(.)$, allows one to set a range of parameters. In order to utilize the same set of range measurements as the one used by our proposed algorithm for this network example, we set the starting condition to random and maximum iterations to 1000. We also choose the stress criterion to be squared and normalized with the sum of 4th powers of the dissimilarities. 

\begin{figure}[!t]
  \centering
  \includegraphics[width=0.45\textwidth]{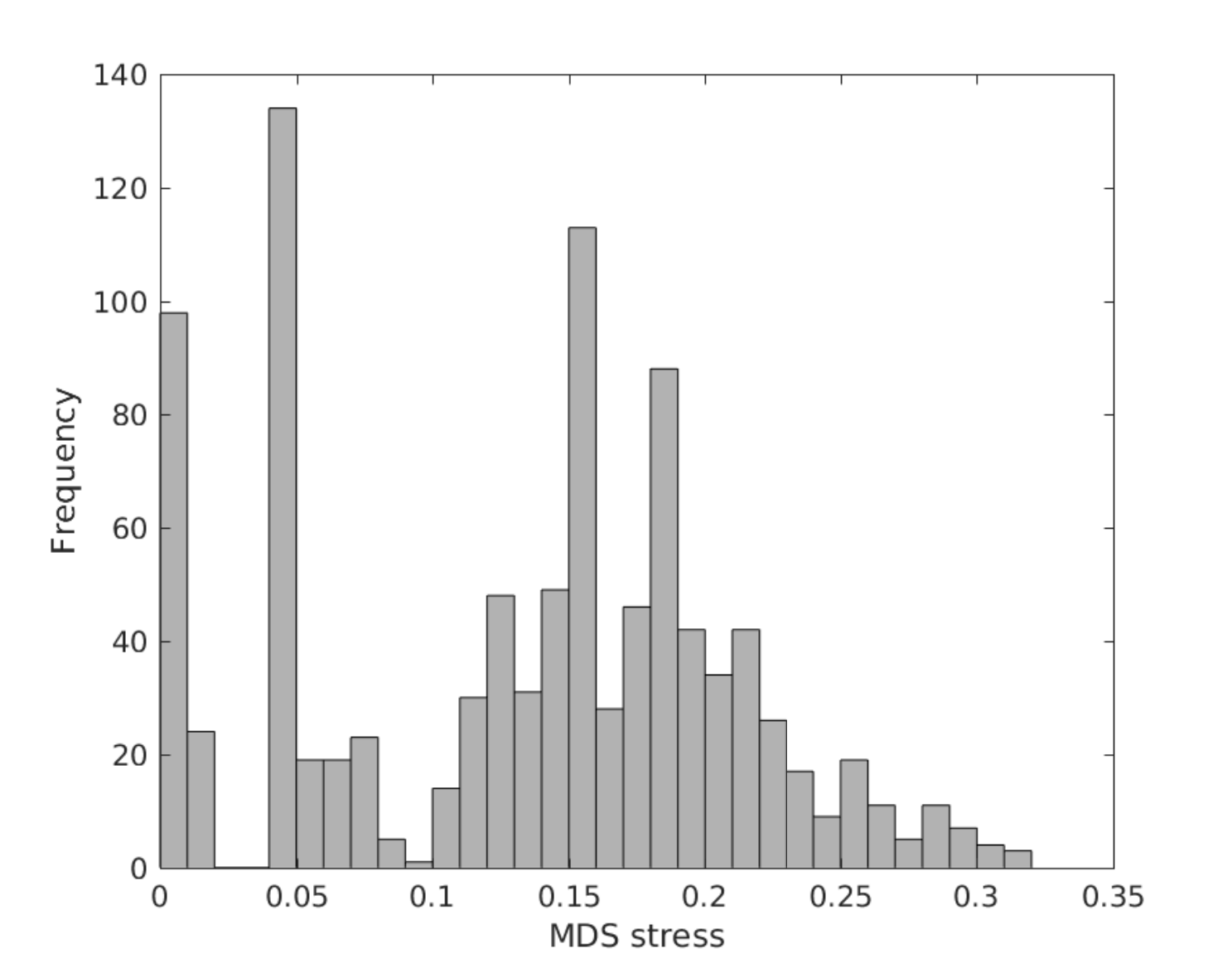}
  \caption{MDS stress value histogram for 1000 simulations with one replication each.}
  \label{fig:MDSStressHistogram}
\end{figure}

It is also possible to choose the number of replications used by the algorithm, i.e. the number of random re-initializations used to compute results, where the one which provides the least stress is chosen as the final answer. In order to compare algorithms, we simulated the MDS algorithm 1000 times with one replication each time. A histogram of MDS stress values is shown in Fig. \ref{fig:MDSStressHistogram}. This histogram was generated using the square root binning method. 

\begin{figure}[!t]
  \centering
  \includegraphics[width=0.45\textwidth]{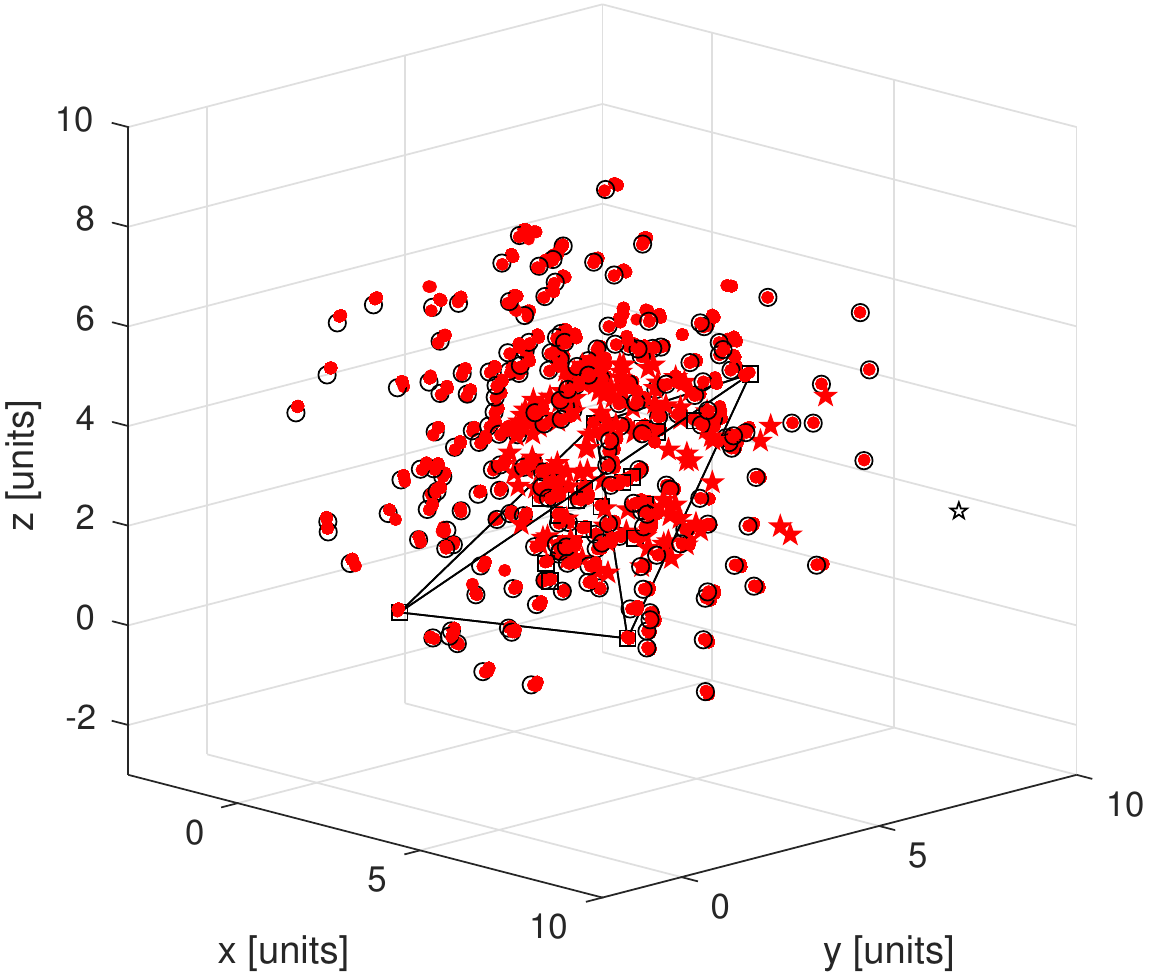}
  \caption{Static sensor network localization example with 216 nodes using all MDS computed node coordinates which have stress value in the most frequent interval from Fig. \ref{fig:MDSStressHistogram}. Given the high number of edges in this network, they are not draw. Instead, we provide a histogram of its node degrees in Fig. \ref{fig:ourNodeDegree}. The black frame are edges of the anchor's convex-hull. Black circles and squares are true node coordinates outside and inside the convex-hull respectively.  Red dots are computed coordinates with the exception of one node which was not localized in any method. This unlocalized node has true coordinates marked by a black star and computed ones marked by red stars.}
  \label{fig:MDSNodeCoordMostCommonStress}
\end{figure}

Fig. \ref{fig:mdsNodeCoord} shows all MDS computed coordinates which have a stress value inside the lowest valued bin interval from the histogram shown in Fig. \ref{fig:MDSStressHistogram}. In this case, except for one node with computed coordinates marked by red stars, all others were correctly localized. Notice that, each red star represent one possible solution in which the MDS stress criterion was less than 0.01, which shows that MDS solutions are not necessarily unique. Moreover, according to the stress value histogram, this best case scenario happened in less than 10\% of all simulations. Lastly, the computed node coordinates for the most frequent stress value interval are shown in Fig. \ref{fig:MDSNodeCoordMostCommonStress}.

\begin{figure}[!t]
  \centering
  \includegraphics[width=0.45\textwidth]{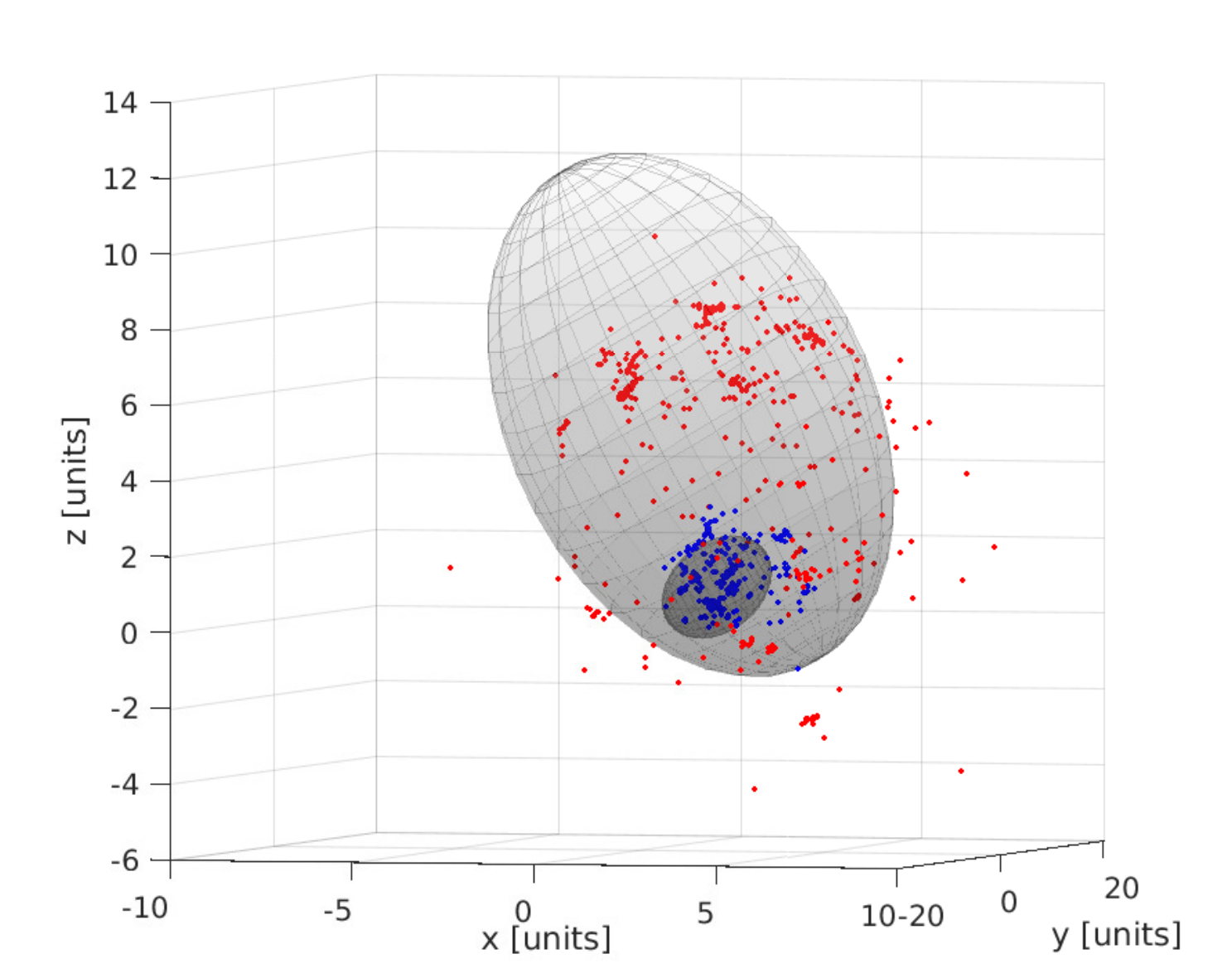}
  \caption{Error ellipsoids for the nodes with computed coordinate covariance matrices that have the smallest and largest maximum eigenvalue. These ellipsoids were obtained so that they encompass 90\% of all 1000 MDS computed coordinates for these nodes, which are also shown as dots in different colors.}
  \label{fig:MDSErrorEllipsoid}
\end{figure}

In order to provide a visual representation of the best and worst computed node coordinates throughout all 1000 simulations of Matlab's MDS, Fig. \ref{fig:MDSErrorEllipsoid}, we use \cite{Johnson} to obtain error ellipsoids which are centered at the average node coordinates. These ellipsoids have axes and orientation proportional to the covariance matrix eigenvalues magnitude and eigenvectors orientation. Their volumes are computed so that 90\% of all computed node coordinates are inside them. These results demonstrate that the MDS algorithm may provide less than desirable results most of the time.

It is interesting to notice that one node, represented by a black star in Fig. \ref{fig:NodeCoord}, was not localized by any method. Besides being outside the anchor's convex-hull, it has less than $n+1$ neighbors given the defined maximum measurement range.

\subsection{Algorithm complexity and execution times}
\label{sec:ComputationComplexity}

An implementation of our algorithm will have worst case scenario for its computational complexity whenever its network graph $\mathcal{G}$ is complete, i.e. all nodes are inter-connected. In which case, the computation of all family sets, $\mathcal{I}_l$ from equation \eqref{eq:FamilySets}, has complexity of $O(m^{n+2})$, where $m$ is the number of nodes and $n$ is the dimension of each node coordinate.

Moreover, there will be $m$ choose $n+1$ sets of $n+1$ neighbors for each node. Therefore, in order to compute matrix $G$ from equation $\eqref{eq:LinearSystem}$, one needs to perform an order of $O(m^{n+3}n^{3})$ operations, considering that determinants of square matrix of dimension $k$ have computational complexity of $O(k^3)$. If one uses Gaussian Elimination or LU-factorization in order to solve the linear system given by equation \eqref{eq:ErrorFreeSolution}, one needs to perform $O((m-(n+1))^3)$. It is usually the case that $m >> n$, so the complexity becomes $O(m^3)$. 

We emphasize that the worst case scenario in which we have a complete graph is not common. Thus, one may define our algorithm's computational complexity considering the maximum node degree, $N = \max{(\{|\mathcal{N}_l|\}_{l = 1,\cdots,m})}$,  where $\mathcal{N}_l$ is specified in equation \eqref{eq:Neighbors}. Then, the construction of matrix $G$ has complexity of $O(m^2 n^3 N^{n+1})$ and the construction of all families of neighbor sets has $O(mN^{n+1})$. Therefore, the complexity is reduced for $m >> N$.

As previously defined, our algorithm has computational cost highly dependent on the number of inter-node range measurements, which are given by each node degree. Fig. \ref{fig:ourNodeDegree} shows a histogram of node degrees for the sensor network in Fig. \ref{fig:NodeCoord}. In which case, the maximum degree is 86 and the minimum is 5, thus much smaller than the worst case scenario.

Notice that the initial process of finding all possible neighbor node combinations for each node can be done only once in the first initialization. If nodes remain static or move inside the range of their neighbors, no future modification are necessary; if not, one may update each node as needed. A similar argument was given in \cite{KhanDiloc}.

\begin{figure}[!t]
\centering
\subfloat[Proportion of correctly localized networks.]{\includegraphics[width=0.4\textwidth]{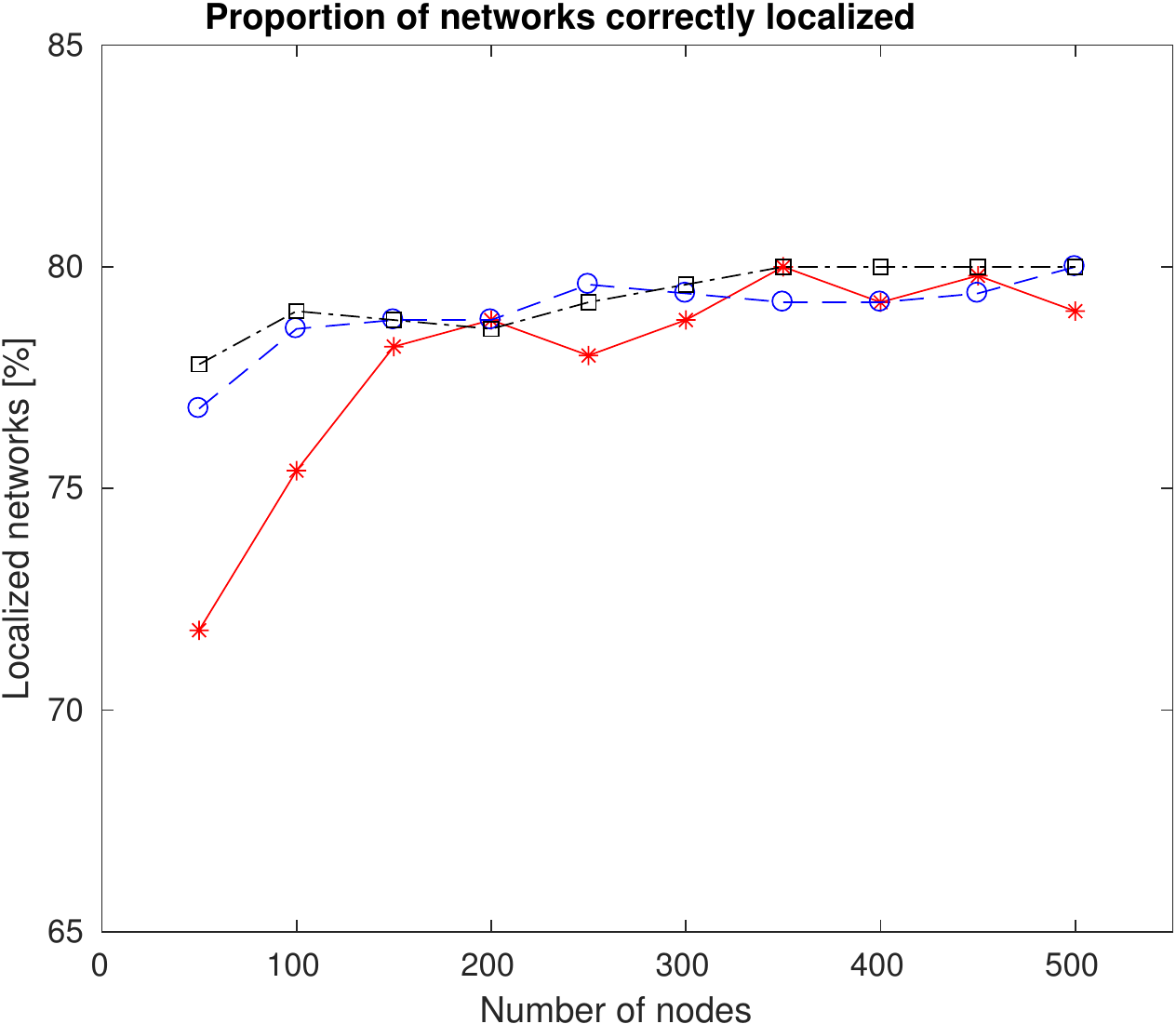}%
\label{fig:BatchProportionCorrectlyLocalized}}
\vfil
\subfloat[Average execution time.]{\includegraphics[width=0.4\textwidth]{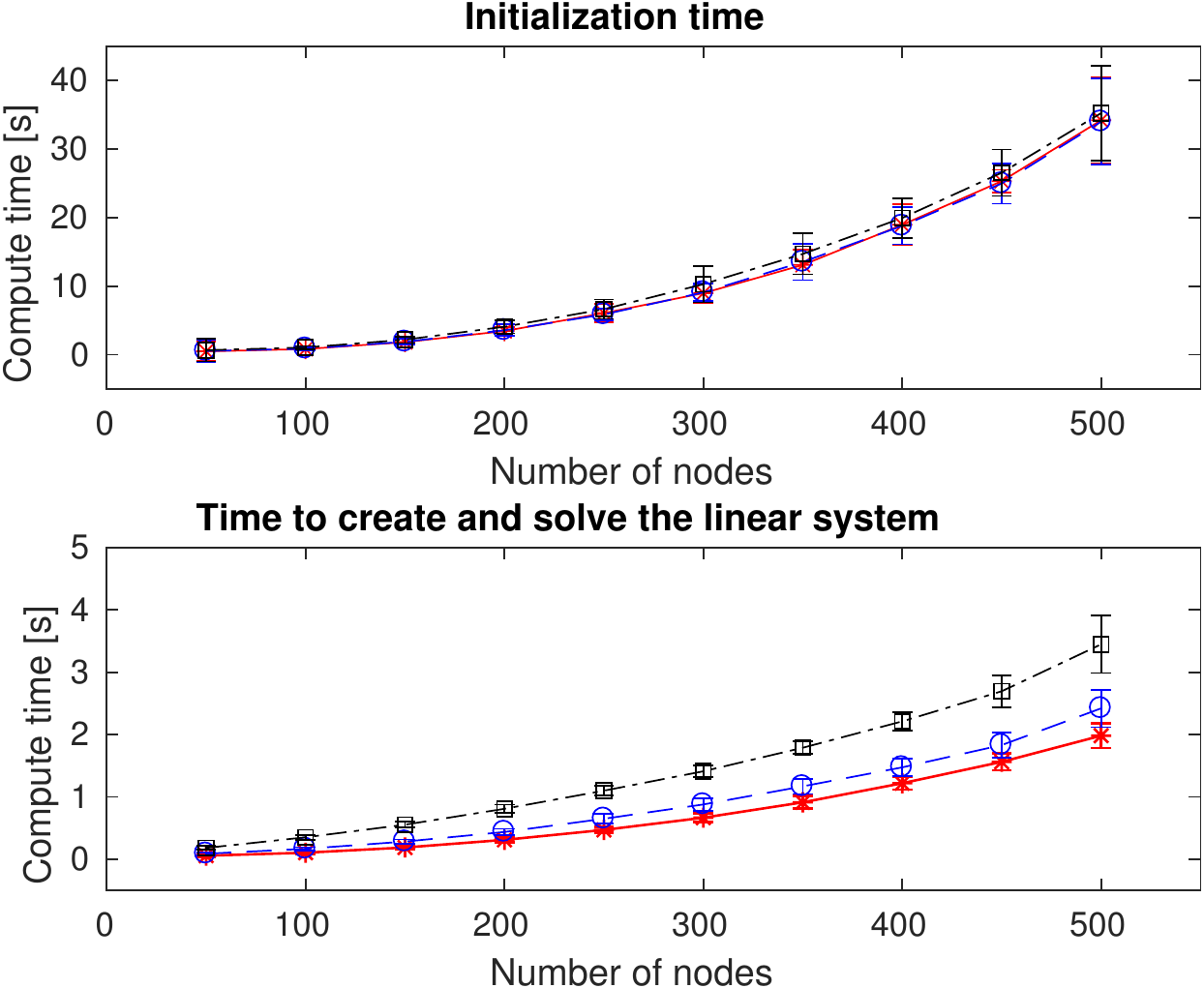}%
\label{fig:BatchAverageExecutionTime}}
\vfil
\subfloat[Average reciprocal condition number.]{\includegraphics[width=0.4\textwidth]{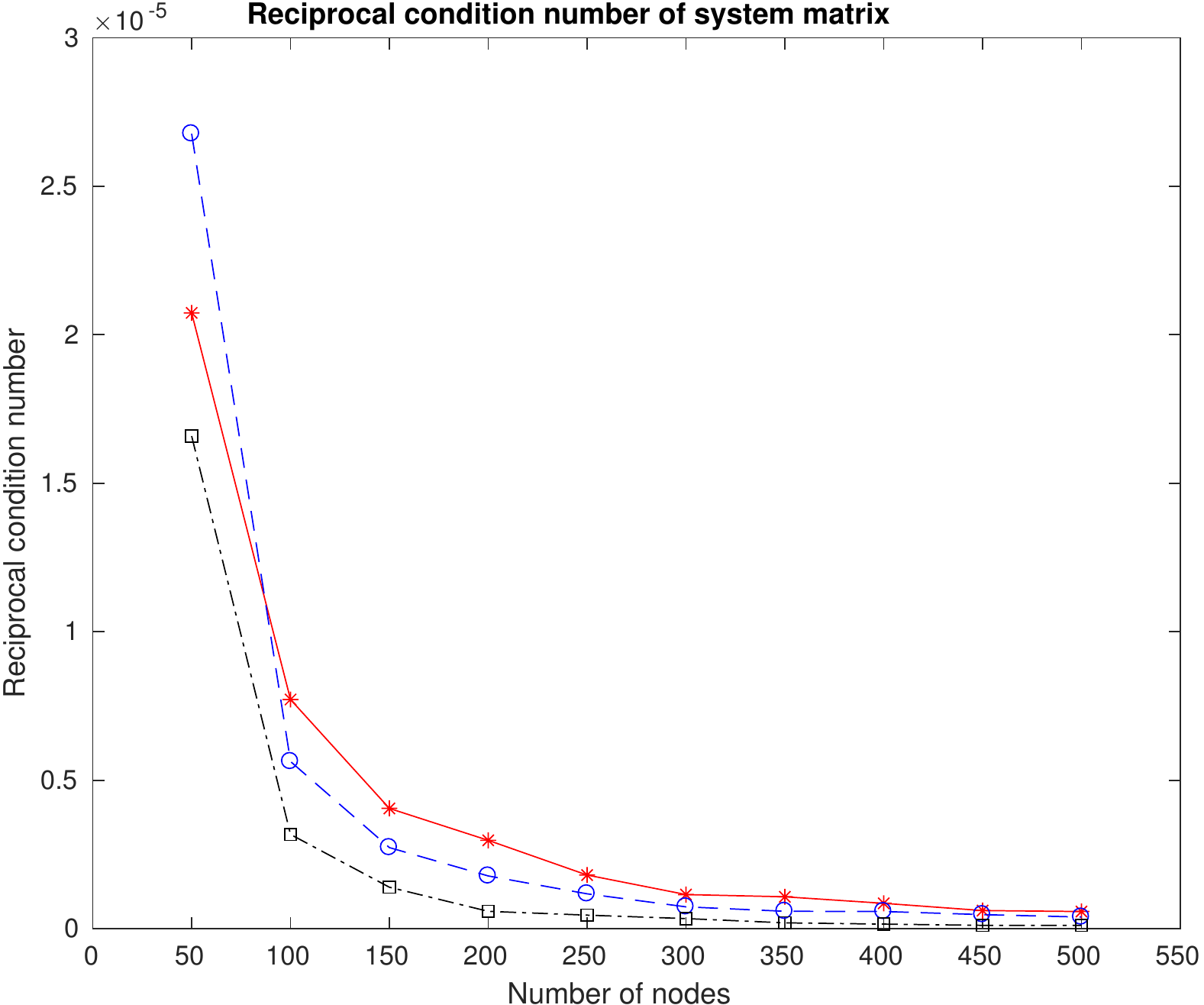}%
\label{fig:BatchAverageReciprocalCN}}
\caption{Simulations of our modified algorithm using 3D networks with 50 to 500 nodes in 50 nodes increments. For each network size, we create 100 sets of random node coordinates with Gaussian distribution $N(0,5I_3)$. For each instance, we select 5 sets of anchor nodes randomly. Distance threshold of 5 units. $\{$Red continuous, Blue dashed, Black dot dashed$\}$  lines are obtained utilizing the minimum between $\{1,50,200\}$ subset of neighbor nodes per pair of neighboring nodes or all possible subsets respectively.}
\label{fig:BatchRandomNetworks}
\end{figure}

In order to provide an idea of our algorithm's run time and compare it to Matlab's MDS implementation, we executed 100 simulations of the former and 1000 for the later. Each MDS simulation was done with one replication only. The average run time for the first part of our algorithm which constructs all possible families of neighbor sets is 140.13 seconds with a standard deviation of 2.71 seconds. The second part which constructs and solves the linear system takes in average 49.15 seconds with a standard deviation of 1.32 seconds. The average execution time for one run of Matlab's MDS implementation with only one replication was 2.80 seconds with standard deviation of 0.71 seconds. 

Although our run times are 68 times larger than MDS' single replication rounds, one will usually need to compute a few rounds of MDS in order to obtain sufficiently small stresses, as can be inferred from Fig. \ref{fig:MDSStressHistogram}. In which, similar or better solutions than the one presented in Fig. \ref{fig:MDSNodeCoordMostCommonStress} happened approximately in 25\% of all MDS simulations made. Therefore, our run times should be around 17 times larger than MDS'.

Despite the lack of optimization of our Matlab implementation, it is obvious that the problem of finding better neighboring node subsets, in which the localization problem can be solved as fast as needed while preserving the same flexibility provided to the placement of anchor nodes, is an interesting one. In this direction, we experiment with our algorithm by choosing smaller number of neighboring subsets for each node. 

Our original method to form all neighbor subsets is a Depth First Search algorithm. We modified it so that it will stop after finding a specified number of subsets, or after finding all possible subsets, whichever occurs first. In the former case, we could have utilized the depth first methodology without modification; but as we previously hypothesized, the existence a minimum of distinct paths from each unknown node to the anchors seems necessary towards the correctly solution of our localization problem. Therefore, we start our search with only one round of Breadth First Search; later, for each previously computed branch, we use Depth First Search to find a specified number of subsets. This approach guarantees that we have at least one feasible subset involving each neighbor node.

In order to test this algorithm modification, we simulate 3D networks with 50 to 500 nodes in 50 nodes increments. For each network size, we generate 100 sets of random node coordinates using Gaussian distribution $N(0,5I_3)$. For each instance, we randomly choose 5 different sets of anchor nodes in order to compute the unknown node coordinates. Lastly, the distance threshold chosen for all instances was equal to 5 units, as this value was approximately equal to the average inter-node distance between all random network nodes generated. 

After each random network is created with its designated number of nodes, we check for nodes with less than $n+1$ neighbors and nodes in which there are no usable neighbor subsets, i.e. neighbors do not form a complete subgraph of $n+1$ nodes or their nodes form a region with zero volume. All nodes that satisfy these conditions are exclude from the initial network which is re-checked until no further nodes are excluded. We emphasize that this simple procedure does \textit{not} guarantee that all unknown nodes have $n+1$ disjoint paths to anchors.

One can infer from Fig. \ref{fig:BatchProportionCorrectlyLocalized}, that our algorithm was able to correctly find the unknown node coordinates in more than 75\% of all random networks tested, utilizing the simple check of nodes connectivity explained above. Moreover, increasing the number of neighbor subsets per node had a positive effect on the smaller networks. We believe that this effect is reduced in the larger simulated networks due to their increased node density per volume.   

Fig. \ref{fig:BatchAverageExecutionTime} coupled with Fig. \ref{fig:BatchProportionCorrectlyLocalized} shows that our experiment achieved a satisfactory result as the total execution times decreased from around 200 seconds to around 10 seconds for networks with less than 250 nodes. Furthermore, even for networks with 500 nodes, our algorithm execution time became less than 40 seconds in average with a success rate between 77\% and 80\%.

Lastly, we show the average reciprocal condition number of each network $I-D$ matrix as given in Theorem \ref{theo:ProblemSolution} in Fig. \ref{fig:BatchAverageReciprocalCN}. This shows that increasing the number of neighbor subsets per node as well as the node density provides a worse conditioned system in general, making it more susceptible to perturbations. 

\subsection{Towards real world deployment}


Real world measurements are subject to many types of interference, be it random like noises or failures resulting in data loss, among others. While we do not provide a method to deal with such interferences, we believe that methodologies similar to the one employed in \cite{DILAND} may provide a solution.

The linear system matrix $G$ utilized in Theorem \ref{theo:ProblemSolution} is constructed utilizing the available range measurements. Moreover, this matrix construction contains all non-linearities inherent on this localization problem. Therefore, measurement noises will affect the value of each element of this matrix in a non-linear form, making the explicit computation of each element's probability density function complex. 

In order to subvert these complications, which also happen to their DILOC algorithm in \cite{KhanDiloc}, Khan et al. propose a new methodology in \cite{DILAND}. Their approach utilize an averaging process of all received measurements in time. Thus, if all measurement noises have zero mean, their average will converge to the noiseless value after sufficient time has passed. So, each new batch of measurements is averaged with all previous ones before being used to compute the necessary barycentric coordinates. A similar process is also applied to each barycentric coordinate. At each step, the newly computed barycentric coordinate is averaged with its previous existing value through the utilization of converging weights. In \cite{DILAND}, it is proved that this process converges to the true barycentric coordinates given sufficient number of iterations in their algorithm based on DILOC.

Besides the dissimilarities existing between our proposed algorithm and DILOC's, we experimented, using simulations, the application of these averaging process to our algorithm. Some simulations would converge to correct solutions, while others wouldn't. We believe that instabilities associated to eigenvalues of the averaged $G$ matrix, the one containing the barycentric coordinates, as previously mentioned in relation to iterative linear methods are responsible for the low reliability of this simple experiment. A better investigation of the underlining process is certainly needed.

\section{Conclusion}
\label{sec:conclusion}

This work's main contribution is a more concise algorithm for network node localization based on barycentric coordinates in $n$-dimensional Euclidean spaces using noiseless range measurements, when unknown nodes are not necessarily confined to the convex-hull of anchor nodes and their neighbors.

Future work will be centered on extending this result in order to solve network node localization with noisy range measurements, as well as, improve its computational efficiency.

\appendices

\section{Cayley-Menger Bi-determinant proof}
\label{app:CayleyMengerBidet}
Following the approach given by \cite{blumenthal1970theory}, we begin by defining matrices $X = [{\bf x_0,x_1,\cdots,x_n}] \in \mathbb{R}^{n\times n+1}$ and $Y = [{\bf y_0,y_1,\cdots y_n}] \in \mathbb{R}^{n\times n+1}$, as

$$
\begin{tabular}{clc}
$ 
V_X =
\begin{bmatrix}
X \\
\mathbf{1}^T
\end{bmatrix}
$ &, and & 
$ 
V_Y =
\begin{bmatrix}
Y \\
\mathbf{1}^T
\end{bmatrix}
$,
\end{tabular}
$$

where $\mathbf{1}$ is the column vector of all ones  with the appropriate dimension.

The signed volume of the sets of points $\mathcal{X}$ and $\mathcal{Y}$, specified by their coordinates given by the columns of matrices $X$ and $Y$, can be found through the determinant of $V_X$ and $V_Y$ respectively:


\begin{equation}
\begin{vmatrix}
V_X
\end{vmatrix}
= (n!) \text{Vol}(\mathcal{X})
\label{eq:detAxVolSX}
\end{equation}
\begin{equation}
\begin{vmatrix}
V_Y
\end{vmatrix}
= (n!) \text{Vol}(\mathcal{Y})
\label{eq:detAyVolSY}
\end{equation}

Next, we can apply a sequence of operations without modifying the value of the determinants of $V_X$ and $V_Y$.
\begin{equation}
\begin{array}{lclcl}
\begin{vmatrix}
V_X
\end{vmatrix} 
& = & 
\begin{vmatrix}
X \\
\mathbf{1}^T
\end{vmatrix}
& = &
\begin{vmatrix}
X & \mathbf{0} \\
\mathbf{1}^T & 0 \\
\mathbf{0}^T & 1
\end{vmatrix} \vspace{2mm}\\
\begin{vmatrix}
V_Y
\end{vmatrix} 
& = & 
\begin{vmatrix}
Y \\
\mathbf{1}^T
\end{vmatrix}
& = &
\begin{vmatrix}
Y & \mathbf{0} \\
\mathbf{1}^T & 0 \\
\mathbf{0}^T & 1
\end{vmatrix}
\end{array}
\end{equation}

For the determinant of $V_X$, we take its transpose and interchange its last two columns and last two rows, obtaining:
\begin{equation}
\begin{array}{lcl}
\begin{vmatrix}
V_X^T
\end{vmatrix}
& = &
\begin{vmatrix}
x_{1 0}   & x_{2 1}   & \hdots & x_{n 0}   & 0 & 1\\
x_{1 1}   & x_{2 1}   & \hdots & x_{n 1}   & 0 & 1\\
\vdots    & \vdots    & \ddots & \vdots    & \vdots & \vdots\\
x_{1 n-1} & x_{2 n-1} & \hdots & x_{n n-1} & 0 & 1\\
0         & 0         & \hdots & 0         & 1 & 0\\
x_{1 n}   & x_{2 n}   & \hdots & x_{n n}   & 0 & 1\\
\end{vmatrix} \vspace{2mm}\\ 
\end{array}
\end{equation}

Multiplying the previous to the determinant of $V_Y$ gives
\begin{equation}
\begin{vmatrix}
V_X^T V_Y
\end{vmatrix}
=
\left|
\begin{smallmatrix}
{\bf x}_0^T{\bf y}_0  & {\bf x}_0^T{\bf y}_1  & \hdots & {\bf x}_0^T{\bf y}_n  & 1\\
{\bf x}_1^T{\bf y}_0  & {\bf x}_1^T{\bf y}_1  & \hdots & {\bf x}_1^T{\bf y}_n  & 1\\
\vdots    & \vdots    & \ddots & \vdots    & \vdots\\
{\bf x}_{n-1}^T{\bf y}_0  & {\bf x}_{n-1}^T{\bf y}_1  & \hdots & {\bf x}_{n-1}^T{\bf y}_n  & 1\\
1         & 1         & \hdots & 1         & 0\\
{\bf x}_n^T{\bf y}_0  & {\bf x}_n^T{\bf y}_1  & \hdots & {\bf x}_n^T{\bf y}_n  & 1\\
\end{smallmatrix}
\right| 
\end{equation}

Interchanging the last two rows returns:
\begin{equation}
\begin{array}{lcl}
\begin{vmatrix}
V_X^T V_Y
\end{vmatrix} \vspace{2mm}
& = &
-
\begin{vmatrix}
{\bf x}_0^T{\bf y}_0  & {\bf x}_0^T{\bf y}_1  & \hdots & {\bf x}_0^T{\bf y}_n  & 1\\
{\bf x}_1^T{\bf y}_0  & {\bf x}_1^T{\bf y}_1  & \hdots & {\bf x}_1^T{\bf y}_n  & 1\\
\vdots    & \vdots    & \ddots & \vdots    & \vdots\\
{\bf x}_n^T{\bf y}_0  & {\bf x}_n^T{\bf y}_1  & \hdots & {\bf x}_n^T{\bf y}_n  & 1\\
1         & 1         & \hdots & 1         & 0\\
\end{vmatrix} \vspace{2mm}\\ 
& = &
-
\begin{vmatrix}
X^TY & \mathbf{1} \\
\mathbf{1}^T & 0
\end{vmatrix} \vspace{2mm} \\
& = &
-
\begin{vmatrix}
M
\end{vmatrix}
\end{array}
\label{eq:detMdetAxTAy}
\end{equation}

Let C be the matrix inside the Cayley-Menger bi-determinant. Using the fact that $d({\bf x}_i, {\bf y}_j)^2 = ||{\bf x}_i - {\bf y}_j||^2 = ({\bf x}_i - {\bf y}_j)^T({\bf x}_i - {\bf y}_j)$, we write
\begin{equation}
\begin{vmatrix}
C
\end{vmatrix}
=
\left|
\begin{smallmatrix}
0 & 1 & \hdots & 1 \vspace{1mm}\\
1 & {\bf x}_0^T{\bf x}_0 + {\bf y}_0^T{\bf y}_0 - 2{\bf x}_0^T{\bf y}_0 & \hdots & {\bf x}_0^T{\bf x}_0 + {\bf y}_n^T{\bf y}_n - 2{\bf x}_0^T{\bf y}_n \vspace{1mm}\\
1 & {\bf x}_1^T{\bf x}_1 + {\bf y}_0^T{\bf y}_0 - 2{\bf x}_1^T{\bf y}_0 & \hdots & {\bf x}_1^T{\bf x}_1 + {\bf y}_n^T{\bf y}_n - 2{\bf x}_1^T{\bf y}_n\\
\vdots & \vdots & \ddots & \vdots\\
1 & {\bf x}_n^T{\bf x}_n + {\bf y}_0^T{\bf y}_0 - 2{\bf x}_n^T{\bf y}_0 & \hdots & {\bf x}_n^T{\bf x}_n + {\bf y}_n^T{\bf y}_n - 2{\bf x}_n^T{\bf y}_n \vspace{1mm}\\
\end{smallmatrix}
\right|
\end{equation}

Applying row and column operations 
$$ Row_i \leftarrow Row_i - {\bf x_{i-2}}^T{\bf x_{i-2}} Row_1 $$ 
$$ Col_j \leftarrow Col_j - {\bf y_{j-2}}^T{\bf x_{j-2}} Col_1 $$
for $ 2 \leq \text{i, j} \leq n+2$, result in:
\begin{equation}
\begin{array}{llll}
\begin{vmatrix}
C
\end{vmatrix}
&=& 
-\frac{1}{2}
&
\begin{vmatrix}
0 & -2 & \hdots & -2 \vspace{1mm}\\
1 & -2{\bf x}_0^T{\bf y}_0 & \hdots & -2{\bf x}_0^T{\bf y}_n \vspace{1mm}\\
1 & -2{\bf x}_1^T{\bf y}_0 & \hdots & -2{\bf x}_1^T{\bf y}_n \vspace{1mm}\\
\vdots & \vdots & \ddots & \vdots \vspace{1mm}\\
1 & -2{\bf x}_n^T{\bf y}_0 & \hdots & -2{\bf x}_n^T{\bf y}_n \vspace{1mm}\\
\end{vmatrix} \vspace{1mm}\\
\end{array}
\end{equation}

Extracting the repeated scalars and noticing that an even number of permutations is required, one can use equation \eqref{eq:detMdetAxTAy}, so that
\begin{equation}
\begin{array}{lcl}
\begin{vmatrix}
C
\end{vmatrix}
& = &
-\frac{(-2)^{n+1}}{2}
\begin{vmatrix}
M
\end{vmatrix} \vspace{2mm}\\
& = &
(-1)^{n+1}2^{n}
\begin{vmatrix}
V_X^T V_Y
\end{vmatrix} \vspace{2mm}\\
\end{array}
\end{equation}
Or, as we defined before,
\begin{equation}
\begin{array}{lcl}
\begin{vmatrix}
V_X^T V_Y
\end{vmatrix}
& = &
2 \left(-\frac{1}{2}\right)^{n+1}
\begin{vmatrix}
C
\end{vmatrix} \vspace{2mm}\\
& = &
D({\bf x}_0, \hdots, {\bf x}_n; {\bf y}_0, \hdots, {\bf y}_n)
\end{array}
\end{equation}

Now, using \eqref{eq:detAxVolSX} and \eqref{eq:detAyVolSY}, we can see that 
\begin{equation}
D({\bf x}_0, \hdots, {\bf x}_n; {\bf y}_0, \hdots, {\bf y}_n) = 
(n!)^2 \text{Vol}(\mathcal{X}) \text{ } \text{Vol}(\mathcal{Y}).
\end{equation}

By taking the sets $\mathcal{Y} = \mathcal{X}$ one finds that 
\begin{equation}
\begin{array}{lcl}
 D({\bf x}_0, \hdots, {\bf x}_n) & = & D({\bf x}_0, \hdots, {\bf x}_n; {\bf x}_0, \hdots, {\bf x}_n)\\
& = & 
(n!)^2 \text{Vol}(\mathcal{X})^2.
\end{array}
\end{equation}

Therefore, the Cayley-Menger Bi-determinant is proportional to the product of the signed volumes of the sets of points as previously defined.

\section*{Acknowledgment}

Pedro P. V. Tecchio would like to thanks CNPq - Brazil for its support.

\ifCLASSOPTIONcaptionsoff
  \newpage
\fi



\bibliographystyle{IEEEtran}
\bibliography{IEEEabrv,ref}

%







\end{document}